\documentclass[11pt,a4paper]{emulateapj}

\usepackage{psfrag}
\usepackage{psfig}
\usepackage{pspicture}
\usepackage{graphicx}

\usepackage{amssymb,amsmath}
\usepackage{natbib}

\def\apj{ApJ}

\submitted{Accepted for publication in the \apj}

\begin{document}

\title {Effects of Local Environment and Stellar Mass on Galaxy Quenching out to \lowercase{$z$} $\sim$ 3}

\author{
Behnam Darvish\altaffilmark{1,2},
Bahram Mobasher\altaffilmark{2},
David Sobral\altaffilmark{3,4,5},
Alessandro Rettura\altaffilmark{6},
Nick Scoville\altaffilmark{1},
Andreas Faisst\altaffilmark{6}, 
and Peter Capak\altaffilmark{6}
}

\setcounter{footnote}{0}
\altaffiltext{1}{Cahill Center for Astrophysics, California Institute of Technology, 1216 East California Boulevard, Pasadena, CA 91125, USA; email: bdarv@caltech.edu}
\altaffiltext{2}{University of California, Riverside, 900 University Ave, Riverside, CA 92521, USA; email: bdarv001@ucr.edu}
\altaffiltext{3}{Department of Physics, Lancaster University, Lancaster, LA1 4YB, UK}
\altaffiltext{4}{Instituto de Astrof\'{\i}sica e Ci\^encias do Espa\c co, Universidade de Lisboa, OAL, Tapada da Ajuda, PT 1349-018 Lisboa, Portugal}
\altaffiltext{5}{Leiden Observatory, Leiden University, P.O. Box 9513, NL-2300 RA Leiden, The Netherlands}
\altaffiltext{6}{Infrared Processing and Analysis Center, California Institute of Technology, Pasadena, CA 91125, USA}
\begin{abstract}

We study the effects of local environment and stellar mass on galaxy properties using a mass complete sample of quiescent and star-forming systems in the COSMOS field at $z\lesssim$ 3. We show that at $z\lesssim$ 1, the median star-formation rate (SFR) and specific SFR (sSFR) of all galaxies depend on environment, but they become independent of environment at $z\gtrsim$ 1. However, we find that only for \textit{star-forming} galaxies, the median SFR and sSFR are similar in different environments, regardless of redshift and stellar mass. We find that the quiescent fraction depends on environment at $z\lesssim$ 1, and on stellar mass out to $z\sim$ 3. We show that at $z\lesssim$ 1, galaxies become quiescent faster in denser environments and that the overall environmental quenching efficiency increases with cosmic time. Environmental and mass quenching processes depend on each other. At $z\lesssim$ 1, denser environments more efficiently quench galaxies with higher masses (log($M/M_{\odot}$)$\gtrsim$ 10.7), possibly due to a higher merger rate of massive galaxies in denser environments, and that mass quenching is more efficient in denser regions. We show that the overall mass quenching efficiency ($\epsilon_{mass}$) for more massive galaxies (log($M/M_{\odot}$)$\gtrsim$ 10.2) rises with cosmic time until $z\sim$ 1 and flattens out since then. However, for less massive galaxies, the rise in $\epsilon_{mass}$ continues to the present time. Our results suggest that environmental quenching is only relevant at $z\lesssim$ 1, likely a fast process, whereas mass quenching is the dominant mechanism at $z\gtrsim$ 1, with a possible stellar feedback physics.
 
\end{abstract}


\section{Introduction} \label{intro1}

Galaxy are distributed into two relatively distinct populations: (1) a population of massive, red, passive galaxies with little to no on-going star-formation activity, mostly occupying what is known as the red sequence, and (2) a population of less-massive, blue, star-forming galaxies that lie on the blue cloud. In the local universe, passive galaxies predominantly inhabit denser environments and galaxy clusters, whereas star-forming systems are mostly found in the field. It has been shown that: (1) the fraction of blue, star-forming galaxies is higher in more distant clusters than those in the low-$z$ and local clusters, an observation known as the Butcher-Oemler effect (e.g., \citealp{Butcher78,Aragon-salamanca93}), (2) the ratio of more massive (giant) to less massive (dwarf) red sequence cluster galaxies decreases with decreasing redshift (e.g., \citealp{Delucia07,Stott07,Foltz15}), (3) there is a significant fraction of post-starburst galaxies (whose star-formation activity has ceased in the past $\sim$ one Gyr prior to the observation) in clusters, higher than in the field at intermediate redshifts (e.g., \citealp{Dressler83,Poggianti09}), and (4) the fraction of S0 galaxies is significantly lower at higher redshift clusters (e.g., \citealp{Dressler99,Postman05}). All these observations suggest that since $z\sim$ 1, a large fraction of star-forming galaxies have truncated their star-formation, and have migrated to the quiescent population that dominates denser environments at lower redshifts, possibly due to the environmental effects.

Several physical quenching processes (with gravitational and/or hydrodynamical origin) that act in denser environments are proposed including ram pressure stripping (e.g., \citealp{Gunn72,Abadi99}), galaxy$-$galaxy interactions (e.g., \citealp{Farouki81,Merritt83}), galaxy harassment (e.g., \citealp{Moore98}), galaxy$-$cluster tidal interactions (e.g., \citealp{Merritt84}), starvation (strangulation) (e.g., \citealp{Larson80,Balogh00}), viscous stripping (e.g., \citealp{Nulsen82}), thermal evaporation (e.g., \citealp{Cowie77}), and halo quenching (e.g., \citealp{Birnboim03}). These mechanisms act at different timescales and physical lengths from the center of the potential well of dense regions, with their strength depending on cosmic time, galaxy properties, and the physics of the quenching environment (for a review on the physics of the environmental mechanisms, see e.g., \citealp{Boselli06}). 

Some of these processes might temporarily enhance star-formation in galaxies due to the compression of the gas (e.g., ram pressure stripping) and/or the inflow of gas into the central part of the galaxy, reviving nuclear activity (e.g., galaxy$-$galaxy interactions; see e.g., \citealp{Mihos96,Kewley06,Sobral15,Stroe15}). However, these mechanisms eventually suppress the star-formation activity by heating and/or removing of gas from galaxies.     
       
However, any galaxy property that depends on the environment also shows some degree of association with stellar mass. For example, on average, more massive galaxies are redder, less star-forming and more likely to have early-type morphologies (e.g., \citealp{Kauffmann04,Baldry06,Fontana09,vandokkum09,Peng10,Nayyeri14}). The internal processes (that are scaled with stellar mass of galaxies) are thought to be associated with e.g., AGN and stellar feedback (e.g., \citealp{Fabian12,Hopkins14}) and can potentially quench galaxies by heating/removal of gas through the deposition of energy and momentum to the interstellar medium (ISM) of galaxies. 

Therefore, two major quenching mechanisms are proposed, that are generally known as ``environmental quenching'' and ``mass quenching'', and they seem to suppress star-formation activity independent of each other (e.g., \citealp{Peng10}). In other words, more massive galaxies are more likely to become quiescent independent of their host environment, and galaxies in denser regions are more likely to become quenched independent of their stellar mass. Moreover, environmental quenching has been attributed to satellites, whereas mass quenching is primarily linked to central galaxies (e.g., \citealp{Peng12}). Physically, it has been proposed that both environment and mass quenching can be explained by the ``halo quenching'' process \citep{Gabor15}, a physical mechanism which states that gas in massive halos (more massive than 10$^{12}$ $M_{\odot}$) is hindered from cooling as it becomes shock-heated \citep{Birnboim03,Dekel06}. 

Despite all the progress, it is still not completely clear how these quenching mechanisms affect properties of galaxies at different redshifts, and their fractional role in suppressing the star-formation as a function of look-back time and galaxy properties. Moreover, due to a possibly strong physical connection and entanglement between internal (stellar mass) and external (environment) processes \citep{DeLucia12,Bolzonella10,Mortlock15,Darvish15a,Davidzon16}, it is not clear why they seem to quench galaxies independent of each other. 

To address these issues, in this paper, we apply the Voronoi tessellation technique to a large mass complete sample of galaxies in the COSMOS field \citep{Scoville07} out to $z\sim$ 3. The Voronoi tessellation method is a robust scale-independent density estimator, it makes no prior assumptions about the size and shape of the physical structures, and it enables us to probe environments in a broad dynamical range, from the dense core of clusters to sparsely populated voids. The COSMOS field is large enough ($\sim$ 2 deg$^{2}$) for the effects of the large-scale structure to be discernible, with a minimal cosmic variance. Our mass complete sample is selected in a consistent manner out to $z\sim$ 3, and is statistically large ($\sim$ 70,000). Using this sample, we explore the role of the local environment of galaxies and their stellar mass on the SFR, sSFR, quiescent fraction, and quiescent fraction growth rate out to $z\sim$ 3. We also study the dependence of environmental effects on stellar mass and galaxy type, the redshift evolution of the trends, and the efficiency of environmental and mass quenching processes.

The format of this paper is as follows. In Section \ref{data}, we briefly review the data, the selection of quiescent and star-forming galaxies and the method used to define their local environment. In Section \ref{science}, we present the main results and discuss them in Section \ref{disc}. A summary of this work is given in Section \ref{sum}.

Throughout this work, we assume a flat $\Lambda$CDM cosmology with $H_{0}$=70 kms$^{-1}$ Mpc$^{-1}$, $\Omega_{m}$=0.3 and $\Omega_{\Lambda}$=0.7. All magnitudes are expressed in the AB system and star-formation rates and stellar masses are based on a Chabrier \citep{Chabrier03} initial mass function (IMF).

\section{Data and Sample Selection} \label{data}

The data used in this study is similar to \cite{Darvish15a}. We use a $K_{s}$-band selected sample of galaxies in the COSMOS field (\citealp{Capak07,Scoville07}; also see \citealp{Laigle16} for the latest data set), with photometric redshifts (photo-$z$) from the COSMOS UltraVISTA catalog \citep{McCracken12,Ilbert13}. Sample galaxies have angular positions in the range 149.3$<\alpha_{2000}$(deg) $<$150.8 and 1.6$<\delta_{2000}$(deg) $<$2.8 ($\sim$ 1.8 deg$^{2}$) and are brighter than $K_{s}<$ 24 (191,151 galaxies). This sample was used to extract the density field of galaxies in a series of overlapping $z$-slices in the COSMOS field, using the weighted version of the Voronoi tessellation method to reliably determine the local environment of galaxies (see \citealp{Darvish15a} and Section \ref{density-est}). Here, we use the overdensity value (defined as the local surface density divided by the mean local surface density) to describe galaxy environment. We discard galaxies that are $<$ 1 Mpc away from the edge of the survey and large masked areas due to their unrealistic, underestimated density value. The magnitude cut $K_{s}<$ 24 used for galaxy selection leads to a sample whose mass completeness is a function of redshift and galaxy type. We use the method explained in \cite{Pozzetti10} (also see \citealp{Ilbert13,Darvish15a}) to estimate the mass completeness limit as a function of redshift and galaxy type. We rely on the mass completeness limit of quiescent galaxies to define six mass complete samples at 0.1$<z<$3.1 (73,481 galaxies). The properties of these samples are given in \cite{Darvish15a} and Table \ref{table-comp}. We also use rest-frame NUV$-r^{+}$ versus $r^{+}-$J color$-$color plot to select quiescent and star-forming galaxies in our mass complete samples. Quiescent galaxies are selected as those with NUV$-r^{+}$ $>$ 3.1 and NUV$-r^{+}$ $>$ 3($r^{+}-$J)+1 \citep{Ilbert13}.  

\section{Methods} \label{method}

In this section, we briefly describe the photo-$z$ accuracy of the data, stellar mass and SFR estimation, and the method used to estimate the density field of galaxies.

\subsection{Photo-z Accuracy}

Since our density estimation method relies on the photo-$z$ of galaxies, accurate and reliable photometric redshift measurements are necessary. Here, we use the photometric redshifts from the COSMOS UltraVISTA catalog \citep{McCracken12,Ilbert13}. The photometric data in 30 bands allow for an accurate measurement of the photo-$z$ up to $z\sim$ 3. A comparison between photometric and spectroscopic redshifts for $K_{s}<$ 24 galaxies shows that photo-$z$ accuracy is $\sigma_{z}$ $\sim$ 0.008 at $z\lesssim$ 1, and it reaches $\sigma_{z}$ $\sim$ 0.03 at $z\sim$ 2-3 \citep{Ilbert13}. Moreover, for the same galaxies, the typical photo-$z$ uncertainties (average of the lower and higher 68\% confidence interval of the photo-$z$ probability distribution function [PDF]) are estimated to be $\Delta z$ $<$ 0.01 at $z$ $<$ 1, reaching $\Delta z \sim$ 0.01 at $z\sim$ 1 \citep{Darvish15a}, consistent with the photo-$z$ vs. spectroscopic redshift comparison. At $z\sim$ 2-3, the typical photo-$z$ PDF uncertainties are $\Delta z$ $\sim$ 0.07-0.1 (larger at $z\sim$ 2), larger than the photo-$z$ vs. spectroscopic redshift uncertainties. However, this is due to the fact that the spectroscopic sample targets brighter galaxies. Indeed, when we limit our sample to brighter galaxies ($K_{s}<$ 22), we find the photo-$z$ PDF uncertainties of $\Delta z$ $\sim$ 0.03-0.05 at $z\sim$ 2-3, more consistent with the uncertainties estimated through the photo-$z$ vs. spectroscopic redshift comparisons.

\begin{table}
\begin{center}
\caption{Properties of Mass Complete Samples} 
\begin{scriptsize}
\centering
\begin{tabular}{lcccc}
\hline
Redshift Range & Mass Completeness Limit & Number of Galaxies \\
               & log($M_{\odot}$) & \\
\hline
0.1$\leq z<$0.5 & 9.14 & 9338\\
0.5$\leq z<$0.8 & 9.47 & 11760\\ 
0.8$\leq z<$1.1 & 9.70 & 13885\\
1.1$\leq z<$1.5 & 9.93 & 13640\\
1.5$\leq z<$2.0 & 9.97 & 12217\\ 
2.0$\leq z<$3.1 & 9.97 & 12641\\ 
\hline
\label{table-comp}
\end{tabular}
\end{scriptsize}
\end{center}

\end{table}

\subsection{Stellar Mass and SFR Estimation} \label{mass-sfr}

Stellar masses and SFRs were obtained by \cite{Ilbert13}, using a SED template fitting procedure to the available UV, optical, and mid-IR photometry. The templates were made using BC03 \citep{Bruzual03}, assuming a Chabrier IMF, three metallicities, an exponentially declining SFH with different timescales ($\tau$ = 0.1-30 Gyr), and the \cite{Calzetti00} extinction law. Contributions from nebular emission lines were considered based on an empirical relation between emission line fluxes and the UV light \citep{Ilbert09}. The typical stellar mass and SFR uncertainties are $\Delta M\sim$ 0.1 dex and $\Delta SFR\sim$ 0.1-0.2 dex, respectively. We also estimated SFRs using the relation between the dust-corrected NUV continuum flux and SFR \citep{Kennicutt98}, and found a very good agreement between the estimated UV- and SED-based SFRs over a broad range of SFRs (-2 $<$ log(SFR)($M_{\odot}$yr$^{-1}$) $<$ 3). The median absolute deviation between the two SFRs is $\sim$ 0.1-0.15 dex. Throughout this work, we only use the SED-based SFRs.

\subsection{Density Estimation} \label{density-est}

The density estimation method has been described in detail in \cite{Darvish15a}. We briefly explain it here. We estimate the density field for a series of overlapping redshift slices ($z$-slice) whose widths are determined by the median of the photo-$z$ uncertainty at each redshift. For each $z$-slice, we assign a weight to each galaxy which shows the likelihood of each galaxy belonging to the $z$-slice of our interest. The weight for each galaxy is determined by measuring the fraction of the photo-$z$ probability distribution function (PDF) of the galaxy that lies within the boundaries of each $z$-slice. The incorporation of the weights is very important when one deals with photo-$z$s to estimate the density field. The weights tend to significantly reduce the projection effect of foreground and background galaxies due to the uncertainties in the photo-$z$s, as only the tail of the photo-$z$ PDF of these contaminants (with small weights) intersects with the $z$-slice of our interest.
 
\cite{Darvish15a} performed two sets of extensive simulations to check the performance of different density estimators, including 5th and 10th nearest neighbors, Delaunay triangulation, adaptive kernel smoothing, and Voronoi tessellation. The simulations showed that the adaptive kernel smoothing and Voronoi tessellation outperform other methods. Here, we use the Voronoi tessellation density estimator. It makes no prior assumptions about the shape of physical structures and it is scale-independent. These characteristics make the Voronoi tessellation immensely robust because in reality, the cosmic web has a complex and multi-scale nature. The Voronoi tessellation estimates the density field over a broad range of scales from core of clusters to sparsely populated voids (e.g., see the range of overdensity values -0.6 $\lesssim$ log(1+$\delta$) $\lesssim$ 1.7 in Section \ref{science}). In order to incorporate the galaxy weights into the Voronoi tessellation method, we used a Monte-Carlo approach that is fully explained in \cite{Darvish15a}.
 
Density field construction based on photometric redshifts automatically suppresses the redshift space distortions that exist in studies based on spectroscopic redshifts because the photo-$z$ uncertainties are typically larger than these distortions. However, very large photo-$z$ uncertainties tend to wash out the structures in the distribution of galaxies. Several studies have shown that using photo-$z$s with typical photo-$z$ uncertainties of $\sigma_{z}$ $\gtrsim$ 0.01 cannot properly reconstruct the density field, especially in very dense regions \citep{Cooper05,Malavasi16}. We note that this is not an issue for our density field estimation out to $z\sim$ 1. However, at higher redshifts, our photo-$z$ uncertainties are large enough that they probably wash out extremely dense environments. That might be partly the reason for the lack of extremely dense regions (log(1+$\delta$)$\gtrsim$ 1.5) beyond $z\gtrsim$ 1.5 in our study (e.g., see Section \ref{science}, Figure \ref{fig:SFR-sSFR}), when combined with the cosmic variance (the size of COSMOS field) and how the large-scale structure grows (i.e., massive halos that are able to effectively quench galaxies are not as numerous at higher redshifts as those at lower redshifts). Nonetheless, even at $z\gtrsim$ 1.5, we can still construct relatively dense environments, and that is because of the incorporation of galaxy weights and the adaptive nature of the $z$-slices (they change according to the typical photo-$z$ uncertainties at each redshift). However, we note that our results at high redshifts might be partly affected by the photo-$z$ uncertainties.  

\section{Results} \label{science}

Using the surface density field constructed in the COSMOS field with the Voronoi tessellation method, we study the dependence of the observable parameters of galaxies, such as SFR and quiescent fraction on their local overdensity and stellar mass. We also investigate the fractional role of stellar mass and environment in quenching the star-formation activity in galaxies.

\subsection{Evolution of SFR and sSFR with Environment, Galaxy Type, and Stellar Mass} \label{SFR-sSFR-env}

In this section, we investigate the dependence of star-formation activity (SFR and sSFR) on local overdensity, stellar mass, and galaxy type and its evolution with redshift. The errorbars incorporate Poisson errors and the cosmic variance uncertainties. The cosmic variance contribution is estimated using \cite{Moster11}. Even for the massive 11$<$log($M/M_{\odot}$)$<$11.5 systems, the fractional uncertainties due to the cosmic variance is small and change between $\sim$ 15-10\% at $z$=0.1-3.1 in the COSMOS field.

\begin{figure*}
 \begin{center}
    \includegraphics[width=7in]{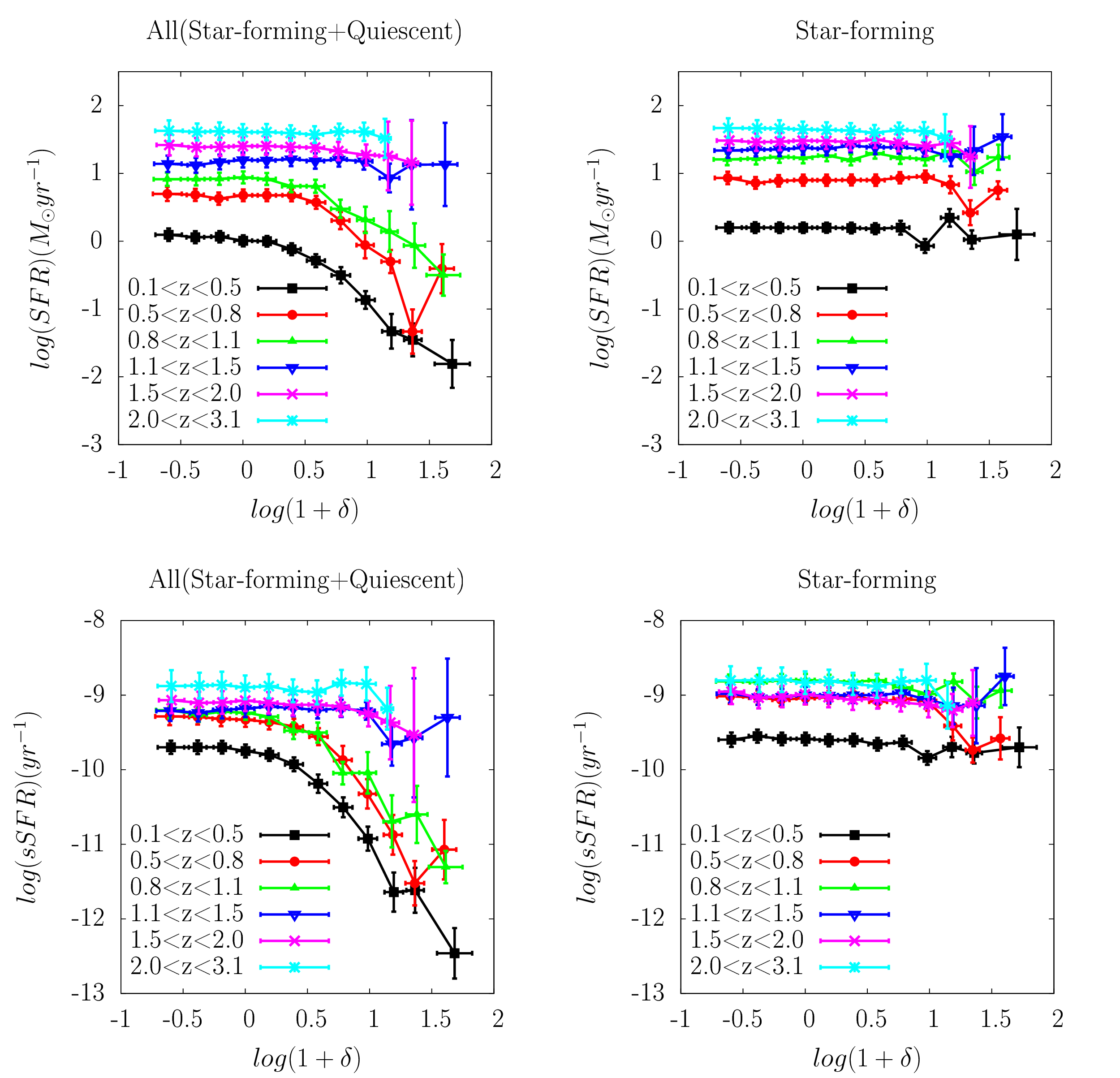}
     \caption{Median SFR (top) and sSFR (bottom) as a function of overdensity at different redshifts for all the galaxies (left) and only the star-forming systems (right). The errorbars incorporate Poisson errors and cosmic variance uncertainties. At $z\lesssim$ 1, the median SFR and sSFR strongly depend on the overdensity and they decrease with increasing overdensity. However, at higher redshifts ($z\gtrsim$ 1), the median SFR and sSFR become almost independent of the environment (overdensity). When we only consider the star-forming galaxies, the median SFR and sSFR become independent of the environment (overdensity) at all the redshifts considered in this study (0.1$<z<$3.1).}
\label{fig:SFR-sSFR}
 \end{center}
\end{figure*}

\subsubsection{Evolution of SFR and sSFR with Environment for the Overall Galaxy Population}

Figure \ref{fig:SFR-sSFR} (top left) shows the median SFR of galaxies (star-forming and quiescent) as a function of overdensity for our mass complete samples at different redshifts. Regardless of the environment, we find that at a fixed overdensity, the median SFR is higher for higher-$z$ samples compared to those at lower redshifts. This is consistent with the continuous decline in the global star-formation density of the universe since $z\sim$ 2$-$3 that has been established by many previous studies (e.g., \citealp{Sobral13,Khostovan15}; see \citealp{Madau14} for a review). 
  
However, we clearly see that the amount of decline depends on the environment. At $z\lesssim$ 1, the median SFR of all the galaxies (quiescent and star-forming) strongly depends on the overdensity and decreases with increasing overdensity, especially for the overdensity values log(1+$\delta$)$\gtrsim$ 0.5. For example, for the 0.1$<z<$0.5 sample, the median SFR decreases by $\sim$ 2 orders of magnitude as the overdensity increases by almost the same factor. However, at higher redshifts ($z\gtrsim$ 1), we do not find a significant relation between the median SFR and overdensity, and the median SFR becomes almost independent of the environment. The decline in the median SFR in low-density environments is $\sim$ 1.6 dex since $z\sim$ 3, whereas this decline in dense environments is $\sim$ 3.6 dex from $z\sim$ 3 to the present time. 

If the environment was mostly relevant for quenching the less massive galaxies, given the fact that our high-$z$ samples do not contain the less massive systems, the environmental independence of the median SFR at $z\gtrsim$ 1 might be due to a selection effect. We investigate this by selecting only galaxies that are more massive than the mass completeness limit of our highest-$z$ sample (log($M/M_{\odot}$)$\geqslant$ 9.97). We find that our results still hold with the new sample selection. The results are presented in the appendix.     

We find similar results for the relation between sSFR and environment for all the galaxies (Figure \ref{fig:SFR-sSFR} (bottom left panel)). At $z\lesssim$ 1, there is a tight anti-correlation between the median sSFR and the overdensity for all the galaxies, especially for log(1+$\delta$)$\gtrsim$ 0.5. However, for the $z\gtrsim$ 1 samples, we find that the sSFR becomes independent of the overdensity values. 

Using a large sample of galaxies in the COSMOS field at $z\lesssim$ 3, \cite{Scoville13} found a strong anti-correlation between the median SFR and environmental density percentile for their low redshift samples. However, at $z\gtrsim$ 1.2, \cite{Scoville13} did not find a strong relation between environment and the median SFR of the observed galaxies, as shown in their figures 15 and 16. They found similar trends between the SF timescale (inverse of the sSFR) and environmental density percentile, in the sense that at $z\lesssim$ 1.2, the SF timescale is larger at higher density percentile levels. These results are fully consistent with what we found in this section, and our results confirm the similar previous studies in the COSMOS field.

\subsubsection{Reversal or No Reversal of SF-Density Relation?}

At low-$z$ ($z\lesssim$ 0.2), many studies show a lower SFR or sSFR in denser regions compared to less-dense, field-like environments for all galaxies (see e.g., \citealp{Balogh04,Kauffmann04,Baldry06}), which are in agreement with our result in this section for our low-$z$ samples (Figure \ref{fig:SFR-sSFR}). 

However, at intermediate redshifts ($z\sim$ 1), there is no consistency on the relation between environment and SFR in galaxies. Some studies retrieve the relations found in the local universe (e.g., \citealp{Patel09,Muzzin12}), some provide evidence for the flattening of the local relations (e.g., \citealp{Grutzbauch11,Scoville13}), and there are even reports of a reversal of the SF$-$density relation (e.g., \citealp{Elbaz07,Cooper08,Welikala16}).

We clearly see no reversal of the SF$-$density relation for our z$\sim$ 1 sample, and our results follow the local universe trends, completely consistent with e.g., \cite{Patel09} and \cite{Muzzin12}. As already noted by \cite{Scoville13}, the claim of the SF$-$density reversal by \cite{Cooper08} is hard to judge according to their Figure 9(b). We argue that the reversal seen in \cite{Elbaz07} is probably due to the cosmic variance and the small dynamical range of the environments considered in their study as they only used the GOODS fields. Furthermore, later studies in the GOODS fields by \cite{Popesso11} attributed this reversal to the active galactic nucleus (AGN) contamination, and \cite{Ziparo14} found no evidence for the reversal of the star-formation activity at $z\sim$ 1 using a similar data set. At $z\sim$ 1, \cite{Sobral11} also found an increase of star-formation activity for star-forming galaxies at intermediate densities, likely related to galaxy groups (and/or filaments; \citealp{Darvish14}). However, they found that this enhancement is accompanied by a decline of star-formation activity in the dense environment and clusters. \cite{Sobral11} argued that this is likely the reason for inconsistencies at $z\sim$ 1, as some studies only reach up to intermediate/group environments, while others only focus in rich clusters. 

At $z\gtrsim$ 1, we still see no dependence of the median SFR, sSFR, and the quiescent fraction (see Section \ref{Q-Fraction}) on the environment. This might be partly due to the larger photo-$z$ uncertainties at higher redshifts and the lack of extremely dense regions in the COSMOS field due to the cosmic variance and how the large-scale structure grows over time, although we can still probe relatively dense environments at $z\gtrsim$ 1 (log(1+$\delta$)$\sim$ 1.4). Moreover, our results represent a median value for SFR, averaged over a variety of environments with similar overdensities but possibly different dynamical states and physics. Nonetheless, the literature results in this redshift range are still controversial as some studies have found a reversal of the SF$-$density relation (e.g., \citealp{Tran10,Santos15}), whereas others have shown that some kind of SF$-$density relation still exists well beyond $z\sim$ 1 (e.g., \citealp{Quadri12,Strazzullo13,Kawinwanichakij14,Newman14,Smail14,Hartley15}; Hung et al. 2016 submitted). Part of this controversy might be due to the interpretation of the results and how the environment is selected and its dynamical state. For example, the existence of a large number of star-forming galaxies in clusters does not necessarily mean that the SF$-$density relation is reversed. In addition, the selection of relaxed, mature clusters (selected through e.g., X-ray emission, an overdensity of red galaxies, and the existence of a red-sequence) automatically biases the environmental study towards a population of passive, quenched galaxies, whereas the selection of less-mature, protoclusters that are still in the formation process (selected through e.g., Ly$\alpha$ or H$\alpha$ overdensities, \citealp{Chiang15}) leads to a bias in favor of blue, star-forming systems.

Therefore, we argue that the cause of discrepancies might be due to cosmic variance, different dynamical range of environments probed, the inclusion of e.g., AGNs in the study, different SFR measures with different timescales, the interpretation of the results, how the environment is selected, and the dynamical state of the environment. Detection of a large sample of structures with different dynamical states in large-volume surveys such as LSST, Euclid, and WFIRST will shed more light on this topic in near future.  
  
\subsubsection{Evolution of SFR and sSFR with Environment and Stellar Mass for Star-forming Galaxies}

Although we find a strong SFR$-$overdensity anti-correlation for the $z\lesssim$ 1 samples for all galaxies, this correlation depends on the galaxy type. Figure \ref{fig:SFR-sSFR} (top right) shows the median SFR of only the star-forming galaxies as a function of overdensity at different redshifts. When we only consider the star-forming galaxies, the median SFR becomes independent of the environment (overdensity) at all the redshifts considered in this study (0.1$<z<$3.1). In other words, while a galaxy is star-forming, on average, its star-formation activity becomes independent of the environment in which it resides (Figure \ref{fig:SFR-sSFR} (top right)). Therefore, the SFR$-$overdensity trends for the $z\lesssim$ 1 samples seen in Figure \ref{fig:SFR-sSFR} (top left) are due to the existence of the quiescent galaxies that populate denser environments. Denser environments increase the likelihood of a galaxy to become quiescent. We also find no relation between the sSFR and the environment (overdensity) for the star-forming galaxies out to $z\sim$ 3 (Figure \ref{fig:SFR-sSFR} (bottom right panel)).

\begin{figure*}
 \begin{center}
    \includegraphics[width=7in, height=6.8in]{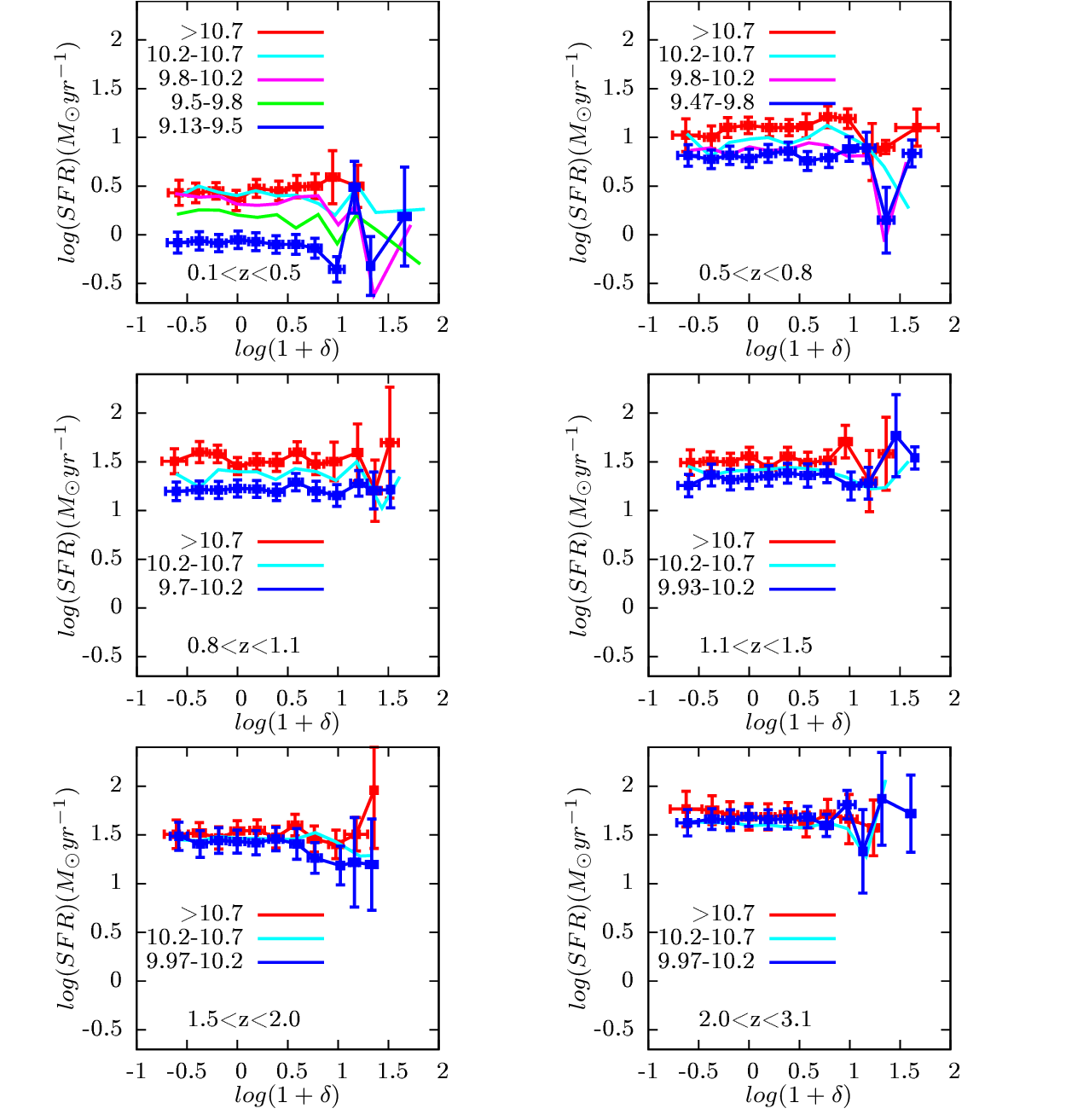}
     \caption{Median SFR as a function of overdensity for different stellar mass bins and at different redshifts for star-forming galaxies only. Different colors correspond to different stellar mass bins. For clarity, the error bars of only the most and the least massive bins are shown. At all the redshifts considered in this study (0.1$<z<$3.1), the median SFR for star-forming galaxies at a fixed environment strongly depends on stellar mass (especially at $z\lesssim$ 1), i.e., the median SFR for star-forming galaxies at all fixed density levels is always higher for more massive galaxies compared to less massive star-forming systems. However, the median SFR for star-forming galaxies at fixed stellar-mass bins is almost independent of their host environment, within the uncertainties. This is a clear representation of the environmental independence of the main-sequence of star-forming galaxies (SFR versus stellar mass), seen in previous studies.}
\label{fig:SFR-env-massbin}
 \end{center}
\end{figure*}

We further divide the star-forming sample into stellar mass bins and investigate the environmental dependence of the median SFR for star-forming systems at fixed stellar mass bins. Figure \ref{fig:SFR-env-massbin} clearly shows that within the uncertainties, the median SFR of star-forming galaxies is also independent of the overdensity at fixed stellar mass bins. This is unexpected as one might expect an environmental dependence of the SFR for (at least) less-massive star-forming galaxies, since several studies at low-$z$ have shown that late-type galaxies in dense environments and clusters are significantly deficient in atomic hydrogen, have lower star-formation activity, and have truncated star-forming and atomic hydrogen disks (for a review, see e.g., \citealp{Boselli06,Boselli14}). We highlight that the environmental dependence of SFR for star-forming galaxies seen in some local universe studies applies mostly to less massive systems ($log(M/M_{\odot}$)$\lesssim$ 9) --- with a mass range not covered in this study --- that are located mostly in dense cluster environments. However, we note that at a given overdensity value, the median SFR for star-forming galaxies is higher for more massive systems out to $z\sim$ 3, especially for lower redshift samples. 

The mass dependence of the SFR for star-forming galaxies is not surprising as many studies have established a tight correlation (typical SFR dispersion of $\sim$ 0.2-0.3 dex) between SFR and stellar mass of star-forming galaxies, that is, the main-sequence of star-forming galaxies (e.g., \citealp{Brinchmann04,Daddi07,Karim11,Sobral14}). We find evidence for the flattening of this relation at higher redshifts (Figure \ref{fig:SFR-env-massbin}). However, the detailed analysis of this is beyond the scope of this work. Moreover, the overdensity independence of the median SFR at fixed stellar mass bins for star-forming galaxies (seen in Figure \ref{fig:SFR-env-massbin}) is another way of presenting the environmental independence of the main-sequence of star-forming galaxies seen in many previous studies (e.g., \citealp{Peng10,Koyama13a,Koyama14,Darvish14}). This indicates that environment does not significantly affect and regulate the mass build-up of the star-forming galaxies through their star-formation activity. However, it decides the probability of a given galaxy to become quiescent.
         
\begin{figure*} 
 \begin{center}
\includegraphics[width=7in]{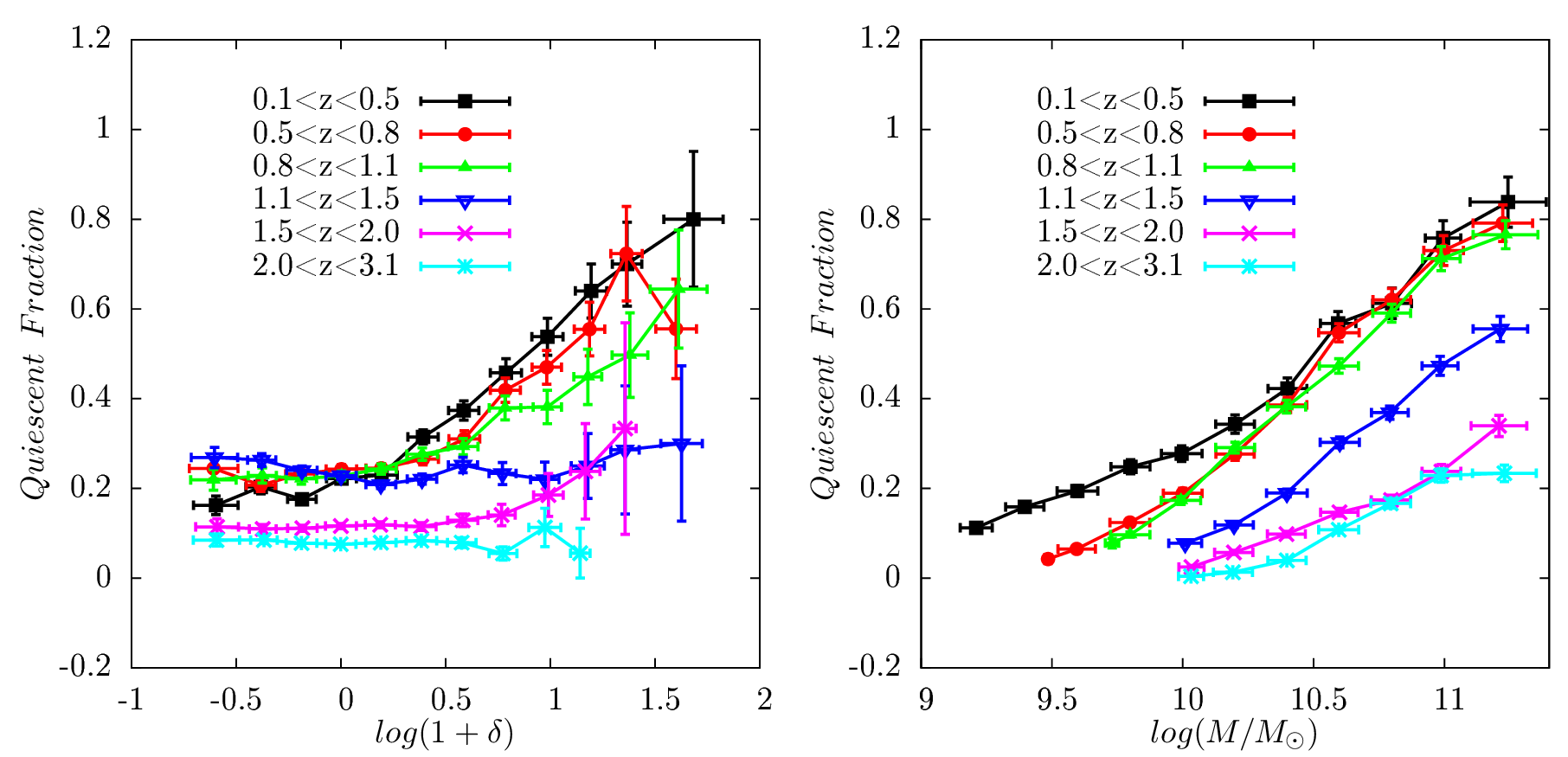}
\caption{Fraction of quiescent galaxies as a function of overdensity (left) and stellar mass (right) and at different redshifts. At $z\lesssim$ 1, we find that the fraction of quiescent galaxies strongly depends on the overdensity. This increases with overdensity from $\sim$ 20\% at log(1+$\delta$)$\lesssim$ 0.5 to $\sim$ 60\%$-$80\% at log(1+$\delta$)$\sim$ 1.6. However, at higher redshifts ($z\gtrsim$ 1), the fraction of quiescent galaxies does not significantly change with environment (overdensity). In addition, we find that at a fixed overdensity, the fraction of quiescent galaxies is higher at lower redshifts, which is a manifestation of the Butcher-Oemler effect. The fraction of quiescent galaxies (located at different environments) strongly depends on stellar mass as well. At all redshifts (for both $z<$1 and $z>$1 samples), the fraction of quiescent galaxies monotonically increases with stellar mass. However, at a fixed stellar mass, this fraction is higher at lower redshifts compared to higher-$z$ samples. This is an indication of the galaxy mass-downsizing.}
\label{fig:fraction-all}
 \end{center}
\end{figure*}

The environmental independence of the median SFR for star-forming galaxies (even at fixed stellar mass bins) agrees well with a plethora of observations and simulations at different redshifts, highlighting that, on average, many properties of the star-forming galaxies that are directly or indirectly linked to star-formation activity (e.g., SFR, sSFR, EW versus stellar mass, main-sequence of star-forming galaxies) do not depend on their host environment (e.g.,  \citealp{Patel09,Peng10,Wijesinghe12,Muzzin12,Koyama13a,Ricciardelli14,Koyama14,Hayashi14,Lin14,
Darvish14,Vogelsberger14,Cen14,Darvish15b,Sobral15,Alpaslan16}; Duivenvoorden et al. 2016 submitted; Hung et al. 2016 submitted).
Indeed, the likelihood of a galaxy to become quenched increases in denser environments. Therefore, many studies note that the main role of environment is to set the fraction of quiescent galaxies (e.g., \citealp{Patel09,Peng10,Sobral11,Muzzin12,Darvish14}; also see Section \ref{Q-Fraction}), whereas for star-forming systems, on average, their star-formation activity is almost independent of the environment. \cite{Darvish15b} suggested the difference in the star-formation histories (SFHs) as a possibility for the difference in the fraction of star-forming galaxies in different environments. 

However, there is no consensus on this topic, as some studies have found an environmental dependence of SFR, even for star-forming galaxies (e.g., \citealp{Vulcani10,Vonderlinden10,Patel11,Haines13,Tran15,Erfanianfar16}). We note that the majority of these studies have found a reduction of $\sim$ 0.1-0.3 dex in the mean SFR of star-forming galaxies in denser environments. This is well within, or of the order of the intrinsic SFR dispersion of $\sim$ 0.2-0.3 dex around the mean main-sequence of star-forming galaxies. \cite{Erfanianfar16} attributed this reduction to a large fraction of red, disk-dominated galaxies in denser regions. \cite{Tran15} found that for a galaxy cluster at $z\sim$ 1.6, the average SFR for H$\alpha$-detected galaxies in the core is half of those in the outer regions. However, their result might have been affected by the incompleteness of their spectroscopic sample in the outer regions. Moreover, the estimated mean sSFR of their sample seems to be independent of the environment. We also note that due to the uncertainties in the SED-based SFRs ($\sim$ 0.1-0.2 dex; see Section \ref{mass-sfr}), in the presence of a weak environmental dependence of the main-sequence, we would not be able to see that in our work. Therefore, we conclude that the inconsistencies in the environmental dependence of the SFR for star-forming galaxies could be due to different selection functions, methods used to define environment \citep{Muldrew12,Darvish15a}, the inclusion of red galaxies in the sample, sample incompleteness, uncertainties in the SFR indicators, etc.

Nevertheless, recent studies have shown that not all characteristics of the star-forming galaxies are independent of environment. For example, gas-phase metallicities were found to slightly depend on environment (e.g., \citealp{Mouhcine07,Shimakawa15}; also see \citealp{Genel16} for detailed simulations), and electron densities and dust extinction were shown to be a strong function of the environment (e.g., \citealp{Koyama13a,Sobral15,Darvish15b,Sobral16}). In the following section, we focus on the fraction of quiescent/star-forming galaxies as a function of overdensity, stellar mass, and redshift and try to disentangle the partial role of the environment and stellar mass on the star-formation quenching in galaxies. 
  
\subsection{Evolution of Quiescent Fraction with Environment and Stellar Mass} \label{Q-Fraction}

In the previous section, we argued that the main role of the environment is to set the fraction of quiescent/star forming galaxies. We now investigate this by studying the fraction of quiescent galaxies as a function of overdensity, stellar mass, and redshift. 

\subsubsection{Quiescent Fraction as a Function of Environment}

\begin{figure*} 
 \begin{center}
\includegraphics[width=7in]{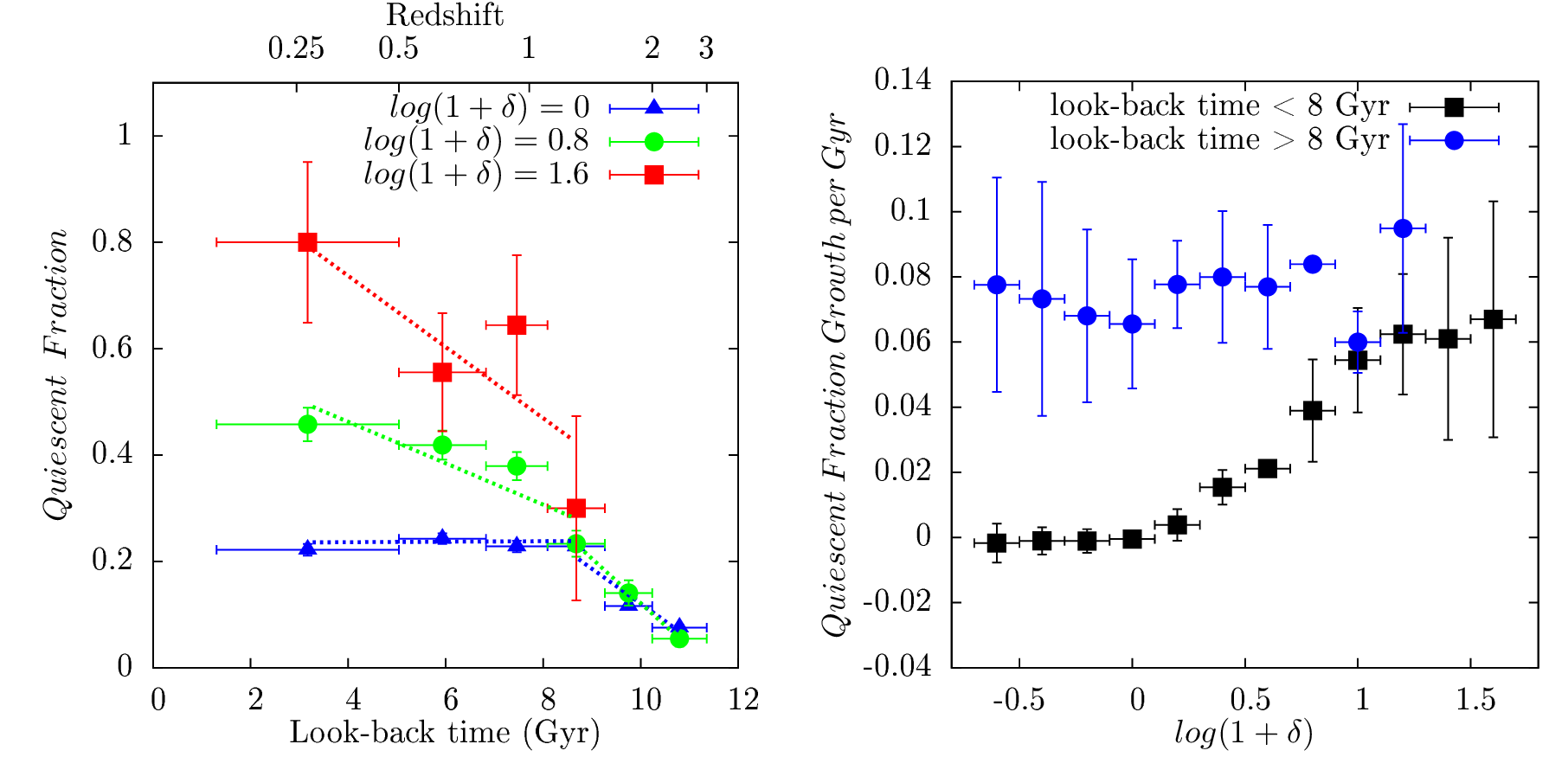}
\caption{(left panel) Fraction of quiescent galaxies as a function of look-back time for some fixed overdensity values. At look-back times $\gtrsim$ 8 Gyr ($z\gtrsim$ 1), the quiescent fraction increases with cosmic time almost independent of the environment. However, at $z\lesssim$ 1 (look-back times $\lesssim$ 8 Gyr), the growth in the fraction of quiescent galaxies with cosmic time is larger for galaxies located in denser environments. (Right panel) Quiescent fraction growth rate (per Gyr) as a function of overdensity for galaxies located at look-back times $\lesssim$ 8 Gyr and $\gtrsim$ 8 Gyr samples. At look-back times $\gtrsim$ 8 Gyr, the quiescent fraction growth rate is almost independent of the environment. However, at look-back times $\lesssim$ 8 Gyr, the growth rate increases with overdensity.}
\label{fig:fraq-grow}
 \end{center}
\end{figure*}

Figure \ref{fig:fraction-all} (left panel) shows the fraction of quiescent galaxies (with different stellar masses) as a function of overdensity and for our mass complete samples at different redshifts. The errorbars are estimated following Poisson statistics and the uncertainties due to the cosmic variance. At $z\lesssim$ 1, we find that the fraction of quiescent galaxies depends on the overdensity. This increases with overdensity from $\sim$ 20\% at log(1+$\delta$)$\lesssim$ 0.5 to $\sim$ 60\%$-$80\% at log(1+$\delta$)$\sim$ 1.6. The increase is specifically steep at log(1+$\delta$)$\gtrsim$ 0.5. However, it does not significantly change with environment (overdensity) at $z\gtrsim$ 1 (except for the 1.5$<z<$2.0 sample at the densest regions, although those have large errorbars). For example, at 2.0$<z<$3.1, the fraction of quiescent galaxies as a function of overdensity remains at $\sim$ 10\% at all overdensity levels. 

Additionally, we find that at a fixed overdensity, the fraction of quiescent galaxies is higher at lower redshifts. This is a clear manifestation of the Butcher-Oemler effect \citep{Butcher78}, stating that the fraction of blue star-forming galaxies is higher for higher-$z$ galaxy clusters compared to their local counterparts (see e.g., \citealp{Butcher78,Couch87,Aragon-salamanca93,Dressler94}). \textit{Moreover, given the environment independence of the median SFR for star-forming galaxies (even at fixed stellar mass bins, Section \ref{SFR-sSFR-env}) and the environmental dependence of their fraction at $z\lesssim$ 1, we conclude that quenching of star-forming galaxies (with stellar masses of log($M/M_{\odot}$)$\gtrsim$ 9 covered in this study) due to environment should happen in a relatively short timescale.}

\subsubsection{Quiescent Fraction Growth Rate}
 
Figure \ref{fig:fraq-grow} (left panel) shows the quiescent fraction as a function of look-back time for some fixed overdensity values. At look-back times $\gtrsim$ 8 Gyr ($z\gtrsim$ 1), the quiescent fraction increases with cosmic time almost independent of the environment. However, we find that at $z\lesssim$ 1 (look-back times $\lesssim$ 8 Gyr) the enhancement in the fraction of quiescent galaxies with cosmic time is larger for galaxies located in denser environments. We quantify the quiescent fraction growth with cosmic time assuming a linear growth in the quiescent fraction at a given overdensity. Figure \ref{fig:fraq-grow} (right panel) shows the quiescent fraction growth rate as a function of overdensity for galaxies located at look-back times $\lesssim$ 8 Gyr and $\gtrsim$ 8 Gyr. At look-back times $\gtrsim$ 8 Gyr, the quiescent fraction growth rate is almost independent of the environment. However, at look-back times $\lesssim$ 8 Gyr and at a given overdensity, the quiescent fraction grows monotonically with cosmic time, with the growth rate increasing with density. Galaxies are transferred from the star-forming population to the quiescent system more rapidly in denser environments compared to less-dense regions. 

Our result regarding the growth of the quiescent fraction agrees qualitatively with \cite{Capak07b} who found a faster growth rate of early-type galaxies (and non-star-forming systems) in denser regions at $z\lesssim$ 1. \cite{Rettura10,Rettura11} also showed that, independent of environment, stellar mass is the main parameter that regulates the formation epoch of early-type galaxies. However, they found that at a given stellar mass, denser environments can trigger a faster mass assembly event for the early-type systems, fully consistent with our result.    

\subsubsection{Quiescent Fraction as a Function of Stellar Mass}

It has been shown that the fraction of quiescent galaxies is also a function of stellar mass (e.g., \citealp{Baldry06,Sobral11}). We investigate the relation between stellar mass and quiescent fraction, as shown in Figure \ref{fig:fraction-all} (right panel). 

According to Figure \ref{fig:fraction-all} (right panel), the fraction of quiescent galaxies (located at different environments) strongly depends on stellar mass. At all redshifts (for both $z<$1 and $z>$1 samples), the fraction of quiescent galaxies monotonically increases with increasing stellar mass. However, at a fixed stellar mass, this fraction is higher at lower redshifts compared to higher-$z$ samples. For example, at 2.0$<z<$3.1, the fraction of quiescent galaxies varies from $\sim$ 0\% for log($M/M_{\odot}$)$\sim$ 10 systems to $\sim$ 20\% for log($M/M_{\odot}$)$\sim$ 11 galaxies. Whereas, at 0.1$<z<$0.5, log($M/M_{\odot}$)$\sim$ 10 galaxies comprise $\sim$ 30\% of quiescent population and this fraction increases to $\sim$ 80\% for log($M/M_{\odot}$)$\sim$ 11 galaxies. 

In fact, this is another form of galaxy ``mass-downsizing'' (e.g., \citealp{Cowie96,Bundy06}), indicating that higher mass galaxies form and quench their star-formation activity earlier than less massive systems. According to Figure \ref{fig:fraction-all} (right panel), by 2.0$<z<$3.1, $\sim$ 20\% of log($M/M_{\odot}$)$\sim$ 11 have already assembled and quenched their star-formation activity, whereas for log($M/M_{\odot}$)$\sim$ 10, this fraction ($\sim$ 20\%) is observed at much later times (at 0.8$<z<$1.1). Using this, we derive an upper limit for the typical time delay between the quenching of less massive (log($M/M_{\odot}$)$\sim$ 10) and more massive (log($M/M_{\odot}$)$\sim$ 11) galaxies and estimate it to be $\Delta t\sim$ 3.4$\pm$0.9 Gyr. 

\begin{figure}
 \begin{center}
    \includegraphics[width=3.5in]{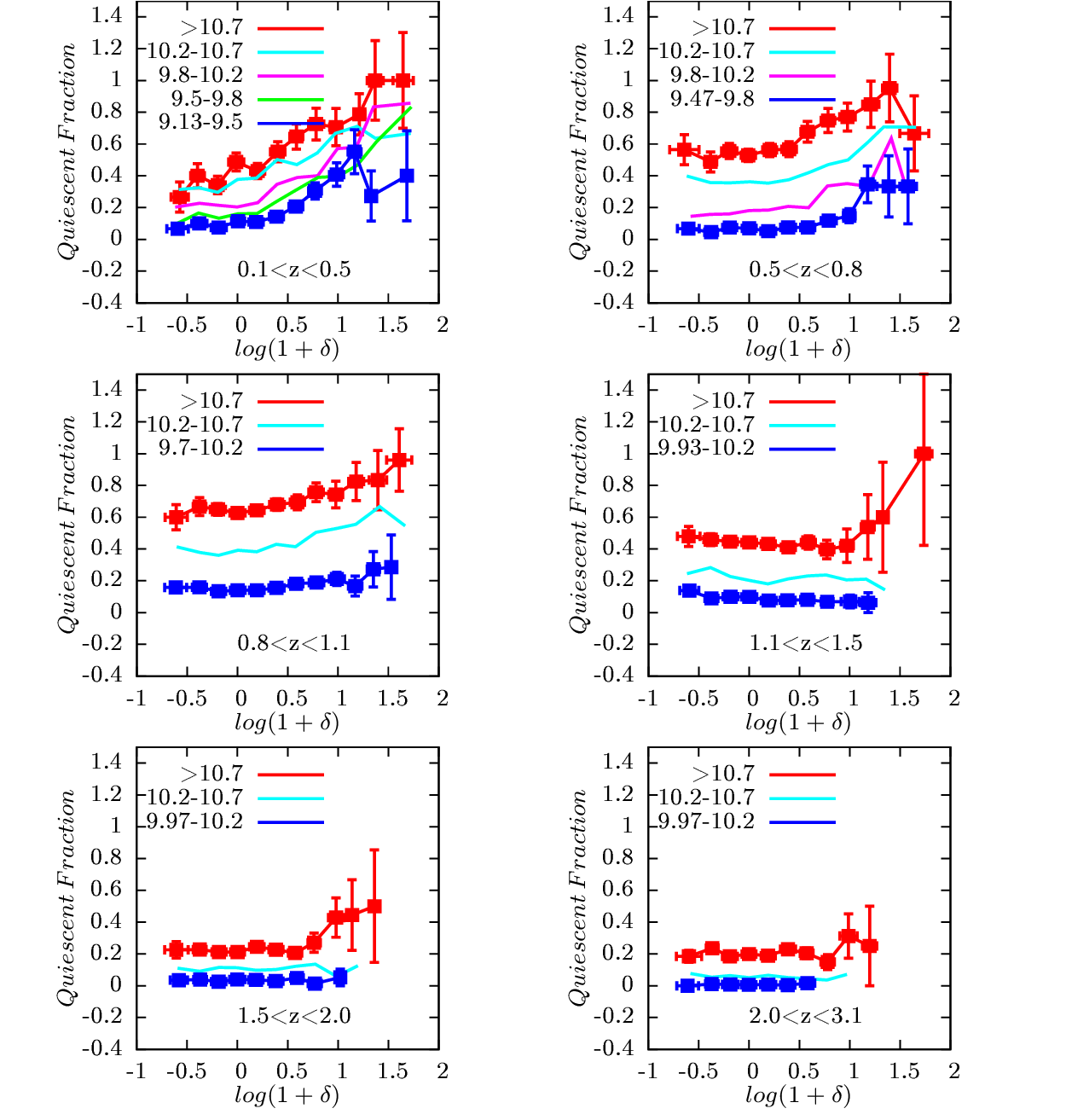}
     \caption{Fraction of quiescent galaxies as a function of overdensity for different stellar mass bins and at different redshifts. Different colors correspond to different stellar mass bins. For clarity, the error bars of only the most and the least massive bins are shown. At all the redshifts considered in this study (0.1$<z<$3.1), the fraction of quiescent galaxies at a fixed environment strongly depends on stellar mass, i.e., the fraction of quiescent galaxies at all fixed density levels is always higher for more massive galaxies compared to less massive systems out to $z\sim$ 3.}
\label{fig:frac-env-mass}
 \end{center}
\end{figure}

\begin{figure}
 \begin{center}
    \includegraphics[width=3.5in]{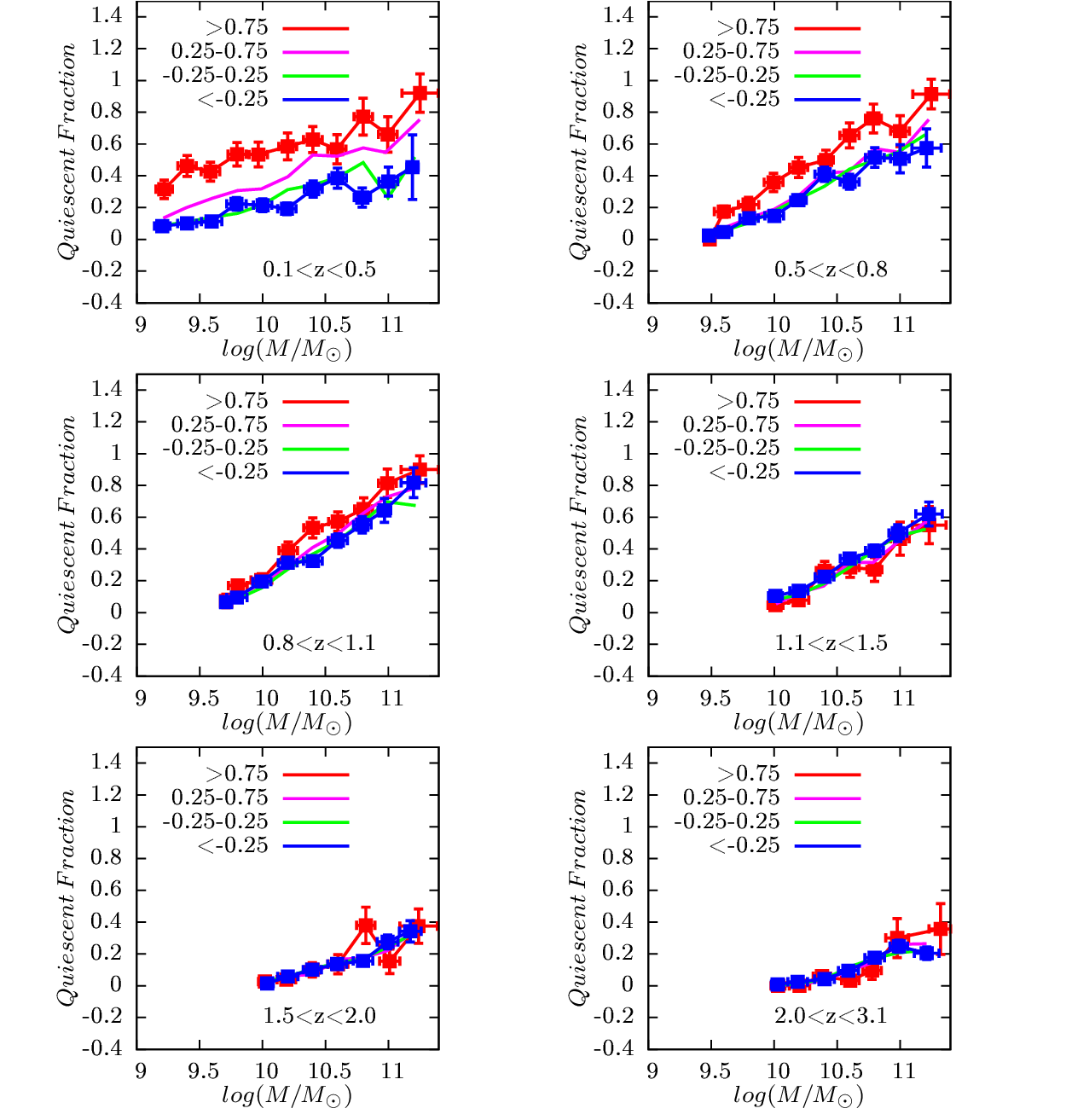}
     \caption{Fraction of quiescent galaxies as a function of stellar mass for different overdensity bins and at different redshifts. Different colors correspond to different overdensity bins. For clarity, the error bars of only the densest and the least-dense overdensity bins are shown. At a fixed stellar mass, quiescent fraction depends on environment (overdensity) only at $z\lesssim$ 1 and not significantly at higher redshifts. At a fixed stellar mass and at z$\lesssim$ 1, this fraction is higher in denser regions.
}
\label{fig:frac-mass-env}
 \end{center}
\end{figure}

\subsubsection{Quiescent Fraction as a Function of Environment and Stellar Mass}

One major difference that is clearly seen in the left and right panels of Figure \ref{fig:fraction-all} is that at all redshifts (0.1$<z<$3.1), the fraction of quiescent galaxies depends on stellar mass, but this fraction depends on overdensity (environment) only up to $z\sim$ 1 and not significantly at higher redshifts. This leads us to the conclusion that environmental quenching is only effective at z$\lesssim$ 1, whereas mass quenching is the dominant quenching mechanism at z$\gtrsim$ 1. 

\begin{figure*}
 \begin{center}
    \includegraphics[width=7in]{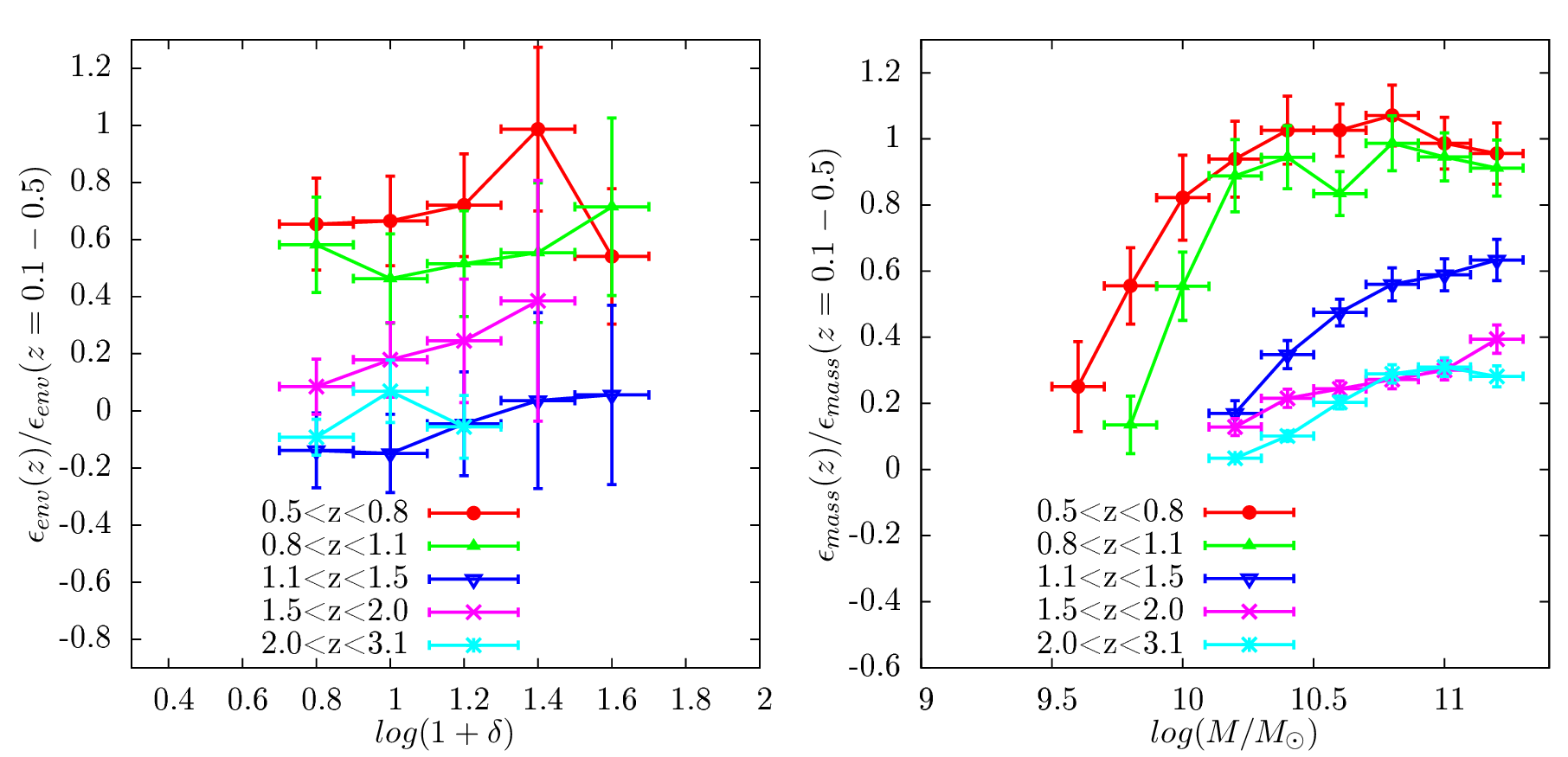}
     \caption{(left panel) Ratio of environmental quenching efficiency at redshift $z$ ($\epsilon_{env}(z)$) to the same quantity for our $z$=0.1-0.5 sample ($\epsilon_{env}(z=0.1-0.5)$), as a function of overdensity values. For clarity, we only show this ratio for denser regions (log(1+$\delta$)$>$0.8). We find that at a given overdensity, $\epsilon_{env}(z)$/$\epsilon_{env}(z=0.1-0.5)$ ratio increases with decreasing redshift, showing that the overall efficiency of the environment in quenching galaxies increases with cosmic time. (right panel) Ratio of stellar mass quenching efficiency at $z$ ($\epsilon_{mass}(z)$) to the same quantity at $z$=0.1-0.5 ($\epsilon_{mass}(z=0.1-0.5)$). For massive galaxies log($M/M_{\odot}$)$\gtrsim$ 10.2), we find that stellar mass quenching efficiency increases with cosmic time from $z\sim$ 3 to $z\sim$ 1. However, at $z\lesssim$ 1, the stellar mass quenching efficiency does not change much with cosmic time for log($M/M_{\odot}$)$\gtrsim$ 10.2 systems. For less massive galaxies (log($M/M_{\odot}$)$\lesssim$ 10.2), the increase in the mass quenching efficiency with cosmic time continues to $z\sim$ 0.1.}
\label{fig:eff}
 \end{center}
\end{figure*}

We further investigate this by studying the fraction of quiescent galaxies as a function of overdensity and redshift for fixed stellar mass bins (Figure \ref{fig:frac-env-mass}) and the fraction of quiescent galaxies as a function of stellar mass and redshift for fixed overdensity bins (Figure \ref{fig:frac-mass-env}). Figure \ref{fig:frac-env-mass} shows the fraction of quiescent galaxies as a function of overdensity for different stellar mass bins and at different redshifts. Different colors correspond to different stellar mass bins. According to Figure \ref{fig:frac-env-mass}, at all the redshifts considered (0.1$<z<$3.1), the fraction of quiescent galaxies at a fixed environment strongly depends on stellar mass, i.e., the fraction of quiescent galaxies at all fixed density levels is always higher for more massive galaxies compared to less massive systems out to z$\sim$ 3. However, as shown in Figure \ref{fig:frac-mass-env}, the fraction of quiescent galaxies at a fixed stellar mass depends on overdensity (shown with different colors) only at $z\lesssim$ 1 and not significantly at higher redshifts. At a fixed stellar mass and at z$\lesssim$ 1, this fraction is higher in denser regions. 

\textit{When we combine the results shown in Figures \ref{fig:fraction-all}, \ref{fig:frac-env-mass} and \ref{fig:frac-mass-env}, we conclude that the quenching of galaxies due to environmental effects (external processes) is effective at lower redshifts ($z\lesssim$ 1), while stellar mass of galaxies (internal processes) is the dominant quenching mechanism at higher redshifts ($z\gtrsim$ 1). We further investigate this in Section \ref{eff-all}.} 

Our results in this section are in general agreement with e.g., \cite{Peng10}, \cite{Sobral11}, \cite{Muzzin12}, and \cite{Lee15}. \cite{Muzzin12} showed that at $z\sim$ 1, the sSFR of star-forming galaxies at fixed environment depends on stellar mass; but at fixed stellar mass, it is independent of environment. \cite{Peng10} studied a sample of galaxies in the SDSS and $z$COSMOS ($z\sim$ 0.7) and showed that for massive galaxies, mass quenching is the primary quenching mechanism for all environments and at all redshifts, while environmental quenching become predominantly important at lower redshifts, especially for less massive galaxies. \cite{Sobral11} showed that stellar mass is the main predictor of star-formation activity in galaxies at $z\sim$ 1, but the environment is responsible for star-formation quenching in all galaxies in dense environments. Recently, \cite{Lee15} found that at $z>$1.4, the quiescent fraction is almost independent of environment, but it is a strong function of stellar mass. They concluded that the stellar mass is the main parameter controlling galaxy quenching out to $z\sim$ 2 for log($M/M_{\odot}$)$>$10 systems, whereas environmental quenching is only relevant at $z<$1. All these results signify the importance of galaxy environment in suppressing the star-formation activity at lower redshifts and stellar mass at higher $z$, and they are qualitatively in agreement with results in this section.

As we already mentioned in Section \ref{density-est}, part of the absence of environmental trends at $z\sim$ 1 might be due to a combination of cosmic variance, how the large-scale structure grows, and the larger photo-$z$ uncertainties at higher redshifts. This sets the need for very large area surveys with large spectroscopic and/or photometric data sets, a possibility that can be achieved in near future with surveys and facilities such as LSST, Euclid, and WFIRST.
    
\subsection{Environmental and Mass Quenching Efficiencies} \label{eff-all}

\begin{figure}
 \begin{center}
    \includegraphics[width=3.5in]{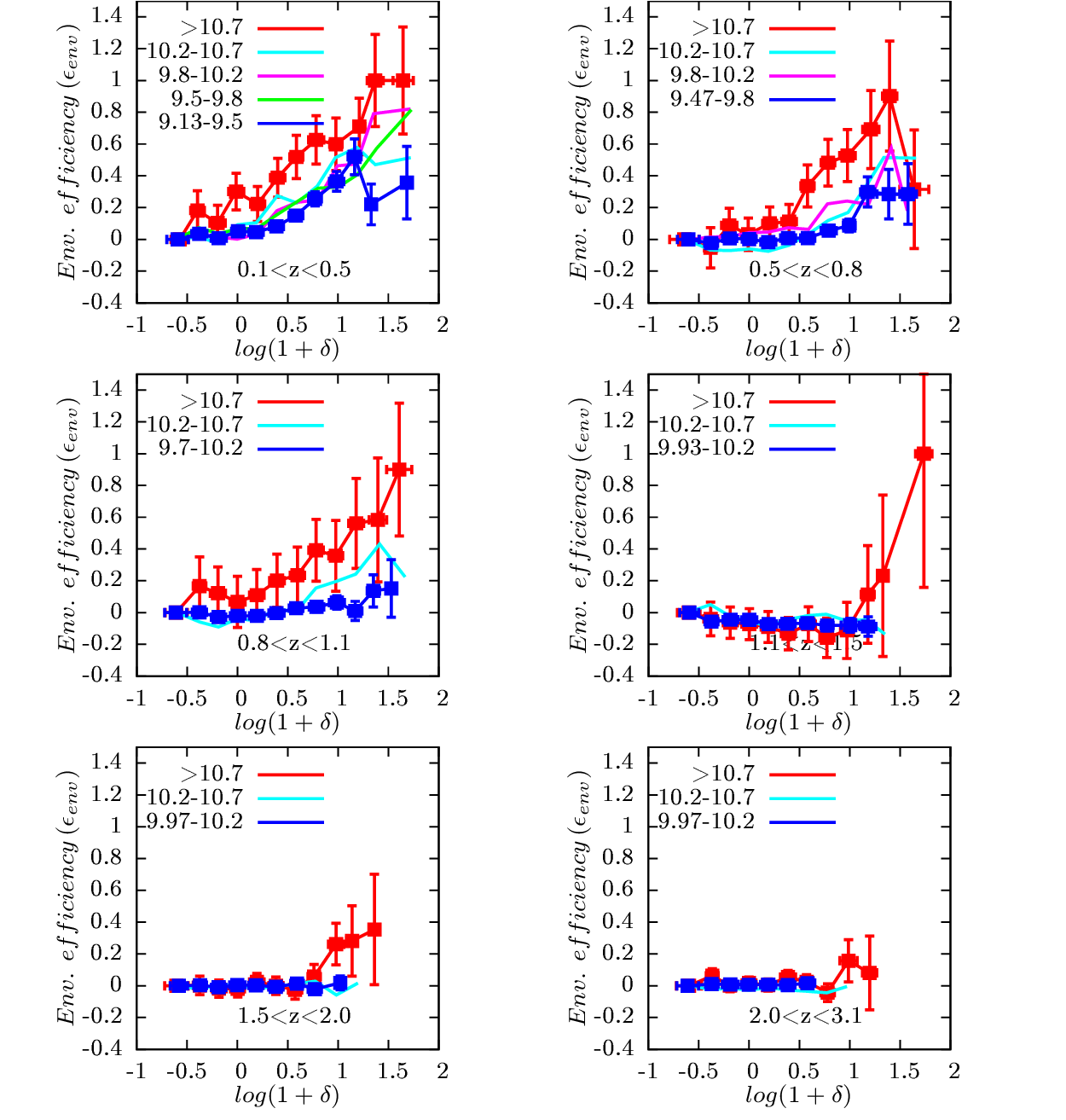}
     \caption{Environmental quenching efficiency as a function of overdensity for different stellar mass bins and at different redshifts. Different colors correspond to different stellar mass bins. For clarity, the error bars of only the most and the least massive bins are shown. Within the uncertainties, the environmental quenching efficiency at a fixed overdensity is almost independent of stellar mass except for very massive system (log($M/M_{\odot}$)$\gtrsim$ 10.7). This mass dependence is more pronounced at lower redshifts and larger overdensity values. This suggests that denser environments (especially at lower redshifts) are able to more efficiently quench galaxies with high stellar masses, possibly due to a higher merger rate of massive systems in denser regions.}
\label{fig:frac-env-eff-mass}
 \end{center}
\end{figure}

\begin{figure}
 \begin{center}
    \includegraphics[width=3.5in]{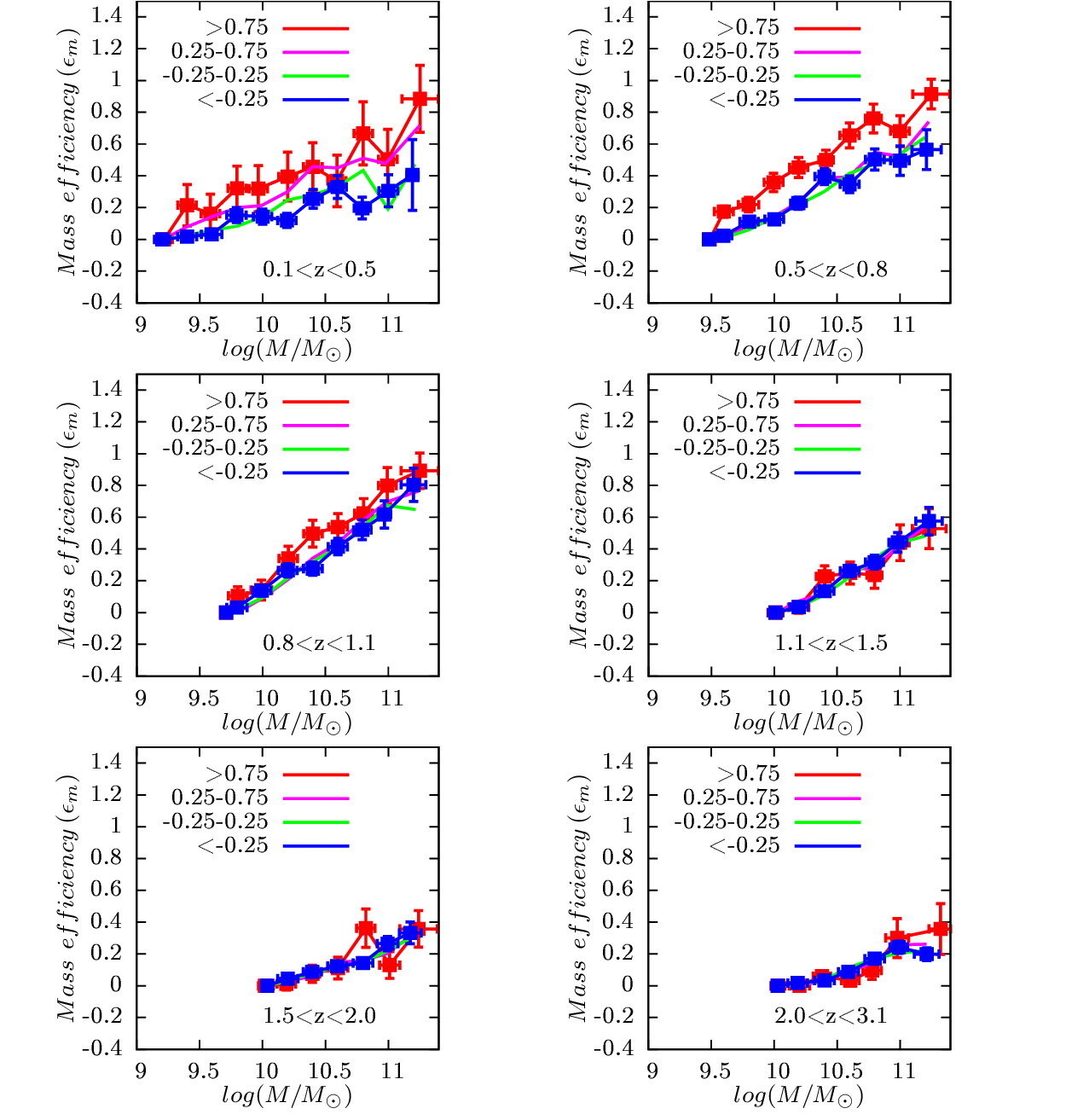}
     \caption{Mass quenching efficiency as a function of stellar mass for different overdensity bins and at different redshifts. Different colors correspond to different overdensity bins. For clarity, the error bars of only the densest and the least-dense overdensity bins are shown. Within the uncertainties, the mass quenching efficiency at a fixed stellar mass is almost independent of overdensity except for very dense environments. This density dependence is more pronounced at lower redshifts and for more massive systems. This suggests that stellar mass quenching acts more efficiently in denser environments (especially at lower redshifts).}
\label{fig:frac-mass-eff-env}
 \end{center}
\end{figure}

In Section \ref{Q-Fraction}, we showed that the quiescent fraction is a function of both environment and stellar mass, in the sense that the quiescent fraction increases with environment and stellar mass. Here, we investigate how efficiently the environment and stellar mass quench galaxies. 

\subsubsection{Definition of Quenching Efficiency}

In order to quantify the ability of environment and stellar mass in quenching the star-formation activity in galaxies, we follow the approach proposed by \cite{Peng10} (also see \citealp{Vandenbosch08,Quadri12,Lin14}). At any given redshift, $z$, we define the \textit{environmental quenching efficiency}, $\epsilon_{env}(\delta,\delta_{0},m,z)$, as the fraction of galaxies at a given stellar mass, $m$, which would be star-forming in low-density environments (with overdensity 1+$\delta_{0}$), but have had their star-formation quenched in denser environments (with overdensity 1+$\delta$) due to some physical processes that are related to environment:
\begin{equation} 
\epsilon_{env}(\delta,\delta_{0},m,z)=\frac{f_{Q}(\delta,m,z)-f_{Q}(\delta_{0},m,z)}{1-f_{Q}(\delta_{0},m,z)}
\end{equation}
 
where $f_{Q}(\delta,m,z)$ is the fraction of quiescent galaxies with stellar mass $m$ that are located in an environment with overdensity 1+$\delta$ at redshift $z$ and $f_{Q}(\delta_{0},m,z)$ is the fraction of quiescent galaxies with stellar mass $m$ that are located in a low-density environment (field) with overdensity 1+$\delta_{0}$ at redshift $z$. In this work, we choose the low-density, field-like environment overdensity as log(1+$\delta_{0}$)$\leqslant$ -0.6. However, we note that fine-tuning around this value does not change the results in this section. Similarly, at any given redshift $z$, we define the \textit{mass quenching efficiency}, $\epsilon_{mass}(m,m_{0},\delta,z)$, as the fraction of galaxies at a given overdensity value, 1+$\delta$, which would be star-forming if they were very low-mass systems (with stellar mass $m_{0}$), but have had their star-formation quenched because of their high stellar mass (with stellar mass $m$) due to some physical processes that are related to stellar mass:
\begin{equation} 
\epsilon_{mass}(m,m_{0},\delta,z)=\frac{f_{Q}(m,\delta,z)-f_{Q}(m_{0},\delta,z)}{1-f_{Q}(m_{0},\delta,z)}
\end{equation}

where $f_{Q}(m,\delta,z)$ is the fraction of quiescent galaxies that are located in the overdensity level 1+$\delta$ and have stellar mass of $m$ at redshift $z$ and $f_{Q}(m_{0},\delta,z)$ is the fraction of quiescent galaxies that are located in the overdensity level 1+$\delta$ and have stellar mass of $m_{0}$ at redshift $z$. In selecting the low stellar mass level ($m_{0}$), we use the stellar mass limit at each redshift.

\subsubsection{Quenching Efficiency as a Function of Environment and Stellar Mass}
 
In general, the environmental and mass quenching efficiencies are functions of both environment and stellar mass. However, we first investigate the environmental quenching efficiency as a function of overdensity only (for all stellar masses) and the mass quenching efficiency as a function of stellar mass only (for all environments). Figure \ref{fig:eff} (left) shows the environmental quenching efficiency at redshift $z$ ($\epsilon_{env}(z)$) normalized to the same quantity for our $z$=0.1-0.5 sample ($\epsilon_{env}(z=0.1-0.5)$), as a function of overdensity values. For clarity, we only show this for denser regions (log(1+$\delta$)$>$0.8) as for less dense regions, the $\epsilon_{env}(z)$ values are around zero and a slight fluctuation around zero results in a large variation in the estimated ratio. We find that at a given overdensity, the $\epsilon_{env}(z)$/$\epsilon_{env}(z=0.1-0.5)$ ratio increases with decreasing redshift, showing that the overall efficiency of the environment in quenching galaxies increases with cosmic time, consistent with \cite{Lee15}. We also investigate a similar ratio for mass quenching efficiency (Figure \ref{fig:eff} (right)). For massive galaxies (log($M/M_{\odot}$)$\gtrsim$ 10.2), we find that the stellar mass quenching efficiency increases with cosmic time from $z\sim$ 3 to $z\sim$ 1 and has flattened out since $z\sim$ 1. However, for less massive galaxies (log($M/M_{\odot}$)$\lesssim$ 10.2), the increase in the mass quenching efficiency continues to the present time. Since the overall environmental quenching efficiency is negligible at $z\gtrsim$ 1, the physical processes related to mass quenching are the main channel for quenching of star-formation at $z\gtrsim$ 1. 

Recently, \cite{Hayward15} presented an analytical model for how the stellar feedback processes drive outflows and regulate star-formation activity in the presence of a turbulent ISM. Their model predicts that in massive galaxies (log($M/M_{\odot}$)$\gtrsim$ 10) at $z\lesssim$ 1, stellar-feedback-driven outflows are suppressed as the gas fraction of such galaxies decreases to values below which the outflow fraction decreases significantly. They also showed that this is not the case for less massive systems as they still have a large gas fraction even at low redshifts. Since outflows are capable of quenching galaxies and they are linked to the mass quenching process, this model might support our result and explain why the mass quenching efficiency for massive $\gtrsim$ 10$^{10}$ $M_{\odot}$ galaxies has not changed much since $z\sim$ 1. Moreover, this also suggests that the physics of the mass quenching process is associated mostly to stellar feedbacks \citep{Hopkins14}. In addition, cold gas should also be hindered from flowing into the galaxy disk (e.g., through heating). Therefore, halo quenching process might also play a role.  
                
Furthermore, we study the mass dependence of $\epsilon_{env}$ and the environmental dependence of $\epsilon_{mass}$. Figure \ref{fig:frac-env-eff-mass} shows the environmental quenching efficiency as a function of environment and for several stellar mass bins at different redshifts. We find that within the uncertainties, $\epsilon_{env}$ is almost independent of stellar mass, except for massive galaxies in dense regions and at $z\lesssim$ 1. That is, denser environments more efficiently quench galaxies that are more massive, especially at lower redshifts. Moreover, according to Figure \ref{fig:frac-mass-eff-env}, we also see that within the uncertainties, $\epsilon_{mass}$ is almost independent of environment, except for galaxies located in very dense environments at $z\lesssim$ 1. This suggests that quenching processes that are associated with higher stellar masses act more efficiently in denser environments.  
 
Our result regarding the mass dependence of the environmental quenching efficiency is consistent with \cite{Lin14} who showed that the environmental quenching efficiency depends on stellar mass out to $z\sim$ 0.8, with $\epsilon_{env}$ greater for more massive galaxies. In agreement with our results, \cite{Knobel15} showed that ``satellite quenching'' (the driver of environmental quenching; \citealp{Peng12}) is almost independent of stellar mass, except for the most massive satellites (log($M/M_{\odot}$) $>$ 11), suggesting some degree of association between mass and environmental quenching. However, several previous studies have suggested that the ability of environment to quench galaxies does not depend on the mass of galaxies and the ability of stellar mass to quench galaxies is independent of where galaxies are located in different environments, at least out to $z\sim$ 2 (e.g., \citealp{Peng10} out to $z\sim$ 1 and \citealp{Quadri12} out to $z\sim$ 2). However, our results show that these fail in very dense environments or for very massive systems. We argue that the disagreement between our result and that of \cite{Quadri12} in very dense environments and for very massive systems might be due to the smaller dynamical range of environments considered in \cite{Quadri12} and the association between massive systems and dense environments. Environments in \cite{Quadri12} only cover up to log(1+$\delta$)$\sim$ 0.6 and in this overdensity regime, we find a good agreement between our result and that of \cite{Quadri12}. However, the environments probed by \cite{Peng10} are similar to those investigated in this study and it is still not clear to us the cause of discrepancy in very dense regions and for massive galaxies. However, it might be due to different selection of star-forming and quiescent galaxies than us, as they only use single color information to separate them.

We note that due to the possible entanglement between very dense environments and very massive galaxies especially at $z\lesssim$ 1 (see e.g., \citealp{Bolzonella10,Mortlock15,Darvish15a}), the higher environmental quenching efficiency for massive systems, and the higher mass quenching efficiency in dense environments might have the same origin that is reflected into two apparently different trends (also see e.g., \citealp{Knobel15,Carollo16}).   
     
\section{Discussion} \label{disc}

We showed that for star-forming galaxies in the mass range of this study (log($M/M_{\odot}$)$\gtrsim$ 9), the median SFR is almost independent of environment out to $z\sim$ 3. However, we showed that the fraction of quiescent galaxies increases with overdensity at $z\lesssim$ 1. The combination of these two suggests that the environmental quenching mechanism should occur in a relatively short timescale. Moreover, we showed that denser environments quench massive galaxies more efficiently and stellar mass quenching is also more efficient in denser environments. Therefore, we should consider a physical mechanism which should be able to best interpret our observations. Ram pressure stripping and (major and minor) mergers are among the environmentally related mechanisms that are capable of quenching galaxies on a relatively short timescale ($\lesssim$1 Gyr). However, ram pressure stripping strongly depends on the mass of the galaxy and is primarily effective in quenching less-massive (log($M/M_{\odot}$)$\lesssim$ 9) systems in dense core of clusters. Therefore, although it is a fast-quenching process, it cannot explain why the environmental quenching efficiency is larger for more massive galaxies. 

Many observations have shown that at least part of the evolution of massive quiescent galaxies is due to mergers. For example, \cite{Skelton09} has argued that dry mergers between galaxies along the red sequence can explain the shallower slope of the bright end of the red sequence. \cite{Bundy09} showed that massive galaxies (log($M/M_{\odot}$)$\gtrsim$ 11) are more likely to host merging companions than less massive (log($M/M_{\odot}$)$\sim$ 10) systems. \cite{VanDerWel09} showed that major merger is the dominant evolutionary agent to produce massive, passive galaxies by analysing the distribution of axial ratio of quiescent galaxies in the local universe. The mass-size evolution of quiescent galaxies (e.g., \citealp{VanDerWel08}) clearly shows that mergers play a role in affecting the evolution of passive, early-type systems. Faisst et al. 2016 in prep. showed that fast, merger-induced quenching can better describe the size evolution of massive log($M/M_{\odot}$)$>$11 quiescent galaxies. Additionally, there is evidence that cluster quiescent systems and those in denser environments are larger than their counterparts in the field and low-density regions (e.g., \citealp{Cooper12a,Lani13,Delaye14}) and that massive brightest cluster galaxies have almost doubled their stellar mass since $z\sim$ 1, possibly due to mergers (e.g., \citealp{Lidman12,Lin13,Burke13,Shankar15}).

Due to large number density of galaxies in dense environments, galaxy merger events are more likely to occur in denser environments compared to the field. This is also supported by the environmental dependence of the mean merger rate of dark matter halos seen in numerical simulations (e.g., \citealp{Fakhouri09}). Moreover, there is evidence for the merger rate being higher for more luminous and more massive galaxies in both simulations and observations (e.g., \citealp{Patton08,Xu12}). Interestingly, \cite{Sobral11} showed that the merger rate is higher for more massive H$\alpha$ star-forming systems located in denser environments at $z\sim$ 0.84. Therefore, the merger scenario (with its short quenching timescale) can qualitatively explain why the environmental quenching efficiency is higher for more massive galaxies in denser environment. 

This picture can also explain why the mass quenching efficiency is larger in denser environments. Prior to mergers, the tidal interactions between galaxies causes the gas in the periphery of the interacting systems to get compressed and funnel towards the central part of the galaxy (e.g., \citealp{Mihos96,Kewley06}). This leads to a rejuvenation of the central regions, followed by a temporary enhancement in star-formation activity, accompanied by the central outflows and feedbacks. Since feedbacks are scaled with mass quenching process, this might explain the higher efficiency of mass quenching mechanism in denser environments.

However, if the physics of the mass quenching is mostly attributed to the AGN activity at $z\lesssim$ 1, given the higher mass quenching efficiency in denser environments at lower redshifts, in principle, there should be a relation between the AGNs and denser environments. Nevertheless, studies at lower redshifts have either found an environmental invariance of the AGN fraction or a lower fraction of AGNs in denser environments (e.g., \citealp{Miller03,Kauffmann04,Best05,Martini06,Popesso06,Sobral16}), with evidence for a higher fraction of AGNs in denser environments only at higher redshifts (e.g., \citealp{Martini13,Donoso14}). This might suggest that the physics of mass quenching at $z\lesssim$ 1 is mostly related to non-AGN feedbacks (possibly stellar feedbacks) and that non-AGN feedback processes might be more effective in denser environments at $z\lesssim$ 1. 
  
As we mentioned before, because of a possible connection between very dense environments and very massive galaxies especially at $z\lesssim$ 1, the higher environmental quenching efficiency for massive systems, and the higher mass quenching efficiency in dense environments found in this work might originate from the same physics, reflected into two apparently different trends.

We also note that there is substantial evidence for central and satellite galaxies following different star-formation histories and having different quenching time-scales (e.g., \citealp{Wetzel13}), in a sense that centrals slowly become quenched in a long timescale ($\gtrsim$ 2 Gyr), whereas quenching of satellites happens gradually in a long timescale after their first infall, followed by a rapid $<$ 1 Gyr quenching \citep{Wetzel13}. Since at a given stellar mass, the fraction of quiescent satellites is higher than the quenched centrals (for stellar masses of $\lesssim$ 10$^{11}$ $M_{\odot}$ at $z$=0 and higher redshifts; see e.g., \citealp{Wetzel13,Knobel13,Hirschmann14,Omand14}), our sample is probably populated by satellite systems and that might be the reason for the overall fast environmental quenching mechanism we found in our work. Therefore, a careful separation of galaxies into central and satellites might reveal different environmental quenching timescales in our work, an opportunity that can be performed in the future.
     
\section{Summary and Conclusion} \label{sum}

We use a mass complete sample of color$-$color selected quiescent and star-forming galaxies in the COSMOS field at $z$=0.1-3.1 to study the effects of stellar mass and local environment of galaxies, estimated via the Voronoi tessellation method, on the star-formation activity of galaxies (e.g., median SFR, sSFR, quiescent fraction, environmental and mass quenching efficiencies). The key results from our work are:
\begin{enumerate}
\item At $z\lesssim$ 1, the median SFR and sSFR of all galaxies and the star-forming fraction decrease with increasing overdensity (environment), and they become almost independent of the environment at $z\gtrsim$ 1. When we only consider the star-forming galaxies, their median SFR and sSFR do not significantly change with environment out to $z\sim$ 3, even at a fixed stellar mass, indicating the environmental independence of the main-sequence of star-forming galaxies. However, denser environments increase the chance of a galaxy to become quiescent. Hence, the role of environment is to control the fraction of star-forming and quiescent galaxies. Given the environment independence of the median SFR for star-forming galaxies (even at fixed stellar mass bins) and the environmental dependence of their fraction at $z\lesssim$ 1, we conclude that environmental quenching of star-forming galaxies should happen in a relatively short timescale ($\lesssim$ 1 Gyr).
\item The quiescent fraction growth rate is almost independent of environment at look-back time $\gtrsim$ 8 Gyr ($z\gtrsim$ 1). However, at lower redshifts ($z\lesssim$ 1), it increases with increasing overdensity. Galaxies become quenched more rapidly in denser environments compared to less-dense regions at low-$z$. 
\item Quiescent fraction increases with stellar mass at all the redshifts considered in this study (0.1$<z<$3.1). At a given stellar mass, the fraction of quiescent galaxies is higher at lower redshifts. This is a manifestation of the galaxy mass-downsizing. We estimate an upper limit for the typical time delay between the quenching of less massive (log($M/M_{\odot}$)$\sim$ 10) and more massive (log($M/M_{\odot}$)$\sim$ 11) galaxies to be $\Delta t\sim$ 3.4$\pm$0.9 Gyr. 
\item Quiescent fraction at a fixed environment strongly depends on stellar mass, and it is always higher for more massive galaxies compared to less massive systems out to $z\sim$ 3. However, at a fixed stellar mass, quiescent fraction depends on environment only at $z\lesssim$ 1 and not significantly at higher redshifts. At a fixed stellar mass and at $z\lesssim$ 1, this fraction is higher in denser regions. This suggests that environmental quenching is practically important at $z\lesssim$ 1, whereas mass quenching is the major factor of star-formation truncation in galaxies at $z\gtrsim$ 1.
\item The overall environmental quenching efficiency increases with cosmic time, in the sense that denser environments at lower redshifts more efficiently quench galaxies compared to the same environments at higher redshifts. The environmental quenching efficiency is almost independent of stellar mass, except for very massive galaxies (log($M/M_{\odot}$)$\gtrsim$ 10.7) at $z\lesssim$ 1. Dense environments at $z\lesssim$ 1 are able to more efficiently quench more massive galaxies than less massive systems. For the purpose of quenching, external processes act more efficiently on more massive galaxies. This might be due to a higher merger rate of massive galaxies in denser environments at $z\lesssim$ 1.
\item For massive galaxies (log($M/M_{\odot}$)$\gtrsim$ 10.2), we find that the overall stellar mass quenching efficiency increases with cosmic time from $z\sim$ 3 to $z\sim$ 1 and flattens out since $z\sim$ 1. however, for less massive galaxies (log($M/M_{\odot}$)$\lesssim$ 10.2), the increase in the mass quenching efficiency continues to the present time. The mass quenching efficiency is almost independent of environment except for very dense environments at lower redshifts. For the purpose of quenching, internal processes act more efficiently in denser environments at $z\lesssim$ 1, with a possible non-AGN (stellar) feedback physics for mass quenching. Due to a correlation between stellar mass and environment especially at lower redshifts, the mass dependence of environmental quenching and the environmental dependence of mass quenching at $z\lesssim$ 1 might originate from the same fundamental physics.

\end{enumerate}

\section*{acknowledgements}

We gratefully thank the anonymous referee for reading the manuscript and providing very useful comments which significantly improved the quality of this work. B.D. acknowledges financial support from NASA through the Astrophysics Data Analysis Program (ADAP), grant number NNX12AE20G. D.S. acknowledges financial support from the Netherlands Organization for Scientific Research (NWO) through a Veni Fellowship, from FCT through an FCT Investigator Starting Grant and Startup Grant (IF/01154/2012/CP0189/CT0010), and from FCT grant PEst-OE/FIS/UI2751/2014. This work extensively uses COSMOS UltraVISTA data which are based on data products from observations made with ESO Telescopes at the La Silla Paranal Observatory under ESO program ID 179.A-2005 and on data products produced by TERAPIX and the Cambridge Astronomy Survey Unit on behalf of the UltraVISTA consortium.

\bibliographystyle{apj} 
\bibliography{references}

\begin{thebibliography}{}
\expandafter\ifx\csname natexlab\endcsname\relax\def\natexlab#1{#1}\fi

\bibitem[{{Abadi} {et~al.}(1999){Abadi}, {Moore}, \& {Bower}}]{Abadi99}
{Abadi}, M.~G., {Moore}, B., \& {Bower}, R.~G. 1999, \mnras, 308, 947

\bibitem[{{Alpaslan} {et~al.}(2016){Alpaslan}, {Grootes}, {Marcum}, {Popescu},
  {Tuffs}, {Bland-Hawthorn}, {Brough}, {Brown}, {Davies}, {Driver}, {Holwerda},
  {Kelvin}, {Lara-L{\'o}pez}, {L{\'o}pez-S{\'a}nchez}, {Loveday}, {Moffett},
  {Taylor}, {Owers}, \& {Robotham}}]{Alpaslan16}
{Alpaslan}, M., {Grootes}, M., {Marcum}, P.~M., {et~al.} 2016, \mnras, 457,
  2287

\bibitem[{{Aragon-Salamanca} {et~al.}(1993){Aragon-Salamanca}, {Ellis},
  {Couch}, \& {Carter}}]{Aragon-salamanca93}
{Aragon-Salamanca}, A., {Ellis}, R.~S., {Couch}, W.~J., \& {Carter}, D. 1993,
  \mnras, 262, 764

\bibitem[{{Baldry} {et~al.}(2006){Baldry}, {Balogh}, {Bower}, {Glazebrook},
  {Nichol}, {Bamford}, \& {Budavari}}]{Baldry06}
{Baldry}, I.~K., {Balogh}, M.~L., {Bower}, R.~G., {et~al.} 2006, \mnras, 373,
  469

\bibitem[{{Balogh} {et~al.}(2004){Balogh}, {Eke}, {Miller}, {Lewis}, {Bower},
  {Couch}, {Nichol}, {Bland-Hawthorn}, {Baldry}, {Baugh}, {Bridges}, {Cannon},
  {Cole}, {Colless}, {Collins}, {Cross}, {Dalton}, {de Propris}, {Driver},
  {Efstathiou}, {Ellis}, {Frenk}, {Glazebrook}, {Gomez}, {Gray}, {Hawkins},
  {Jackson}, {Lahav}, {Lumsden}, {Maddox}, {Madgwick}, {Norberg}, {Peacock},
  {Percival}, {Peterson}, {Sutherland}, \& {Taylor}}]{Balogh04}
{Balogh}, M., {Eke}, V., {Miller}, C., {et~al.} 2004, \mnras, 348, 1355

\bibitem[{{Balogh} {et~al.}(2000){Balogh}, {Navarro}, \& {Morris}}]{Balogh00}
{Balogh}, M.~L., {Navarro}, J.~F., \& {Morris}, S.~L. 2000, \apj, 540, 113

\bibitem[{{Best} {et~al.}(2005){Best}, {Kauffmann}, {Heckman}, {Brinchmann},
  {Charlot}, {Ivezi{\'c}}, \& {White}}]{Best05}
{Best}, P.~N., {Kauffmann}, G., {Heckman}, T.~M., {et~al.} 2005, \mnras, 362,
  25

\bibitem[{{Birnboim} \& {Dekel}(2003)}]{Birnboim03}
{Birnboim}, Y., \& {Dekel}, A. 2003, \mnras, 345, 349

\bibitem[{{Bolzonella} {et~al.}(2010){Bolzonella}, {Kova{\v c}}, {Pozzetti},
  {Zucca}, {Cucciati}, {Lilly}, {Peng}, {Iovino}, {Zamorani}, {Vergani},
  {Tasca}, {Lamareille}, {Oesch}, {Caputi}, {Kampczyk}, {Bardelli}, {Maier},
  {Abbas}, {Knobel}, {Scodeggio}, {Carollo}, {Contini}, {Kneib}, {Le
  F{\`e}vre}, {Mainieri}, {Renzini}, {Bongiorno}, {Coppa}, {de la Torre}, {de
  Ravel}, {Franzetti}, {Garilli}, {Le Borgne}, {Le Brun}, {Mignoli},
  {Pell{\'o}}, {Perez-Montero}, {Ricciardelli}, {Silverman}, {Tanaka},
  {Tresse}, {Bottini}, {Cappi}, {Cassata}, {Cimatti}, {Guzzo}, {Koekemoer},
  {Leauthaud}, {Maccagni}, {Marinoni}, {McCracken}, {Memeo}, {Meneux},
  {Porciani}, {Scaramella}, {Aussel}, {Capak}, {Halliday}, {Ilbert},
  {Kartaltepe}, {Salvato}, {Sanders}, {Scarlata}, {Scoville}, {Taniguchi}, \&
  {Thompson}}]{Bolzonella10}
{Bolzonella}, M., {Kova{\v c}}, K., {Pozzetti}, L., {et~al.} 2010, \aap, 524,
  A76

\bibitem[{{Boselli} \& {Gavazzi}(2006)}]{Boselli06}
{Boselli}, A., \& {Gavazzi}, G. 2006, \pasp, 118, 517

\bibitem[{{Boselli} \& {Gavazzi}(2014)}]{Boselli14}
---. 2014, \aapr, 22, 74

\bibitem[{{Brinchmann} {et~al.}(2004){Brinchmann}, {Charlot}, {White},
  {Tremonti}, {Kauffmann}, {Heckman}, \& {Brinkmann}}]{Brinchmann04}
{Brinchmann}, J., {Charlot}, S., {White}, S.~D.~M., {et~al.} 2004, \mnras, 351,
  1151

\bibitem[{{Bruzual} \& {Charlot}(2003)}]{Bruzual03}
{Bruzual}, G., \& {Charlot}, S. 2003, \mnras, 344, 1000

\bibitem[{{Bundy} {et~al.}(2009){Bundy}, {Fukugita}, {Ellis}, {Targett},
  {Belli}, \& {Kodama}}]{Bundy09}
{Bundy}, K., {Fukugita}, M., {Ellis}, R.~S., {et~al.} 2009, \apj, 697, 1369

\bibitem[{{Bundy} {et~al.}(2006){Bundy}, {Ellis}, {Conselice}, {Taylor},
  {Cooper}, {Willmer}, {Weiner}, {Coil}, {Noeske}, \& {Eisenhardt}}]{Bundy06}
{Bundy}, K., {Ellis}, R.~S., {Conselice}, C.~J., {et~al.} 2006, \apj, 651, 120

\bibitem[{{Burke} \& {Collins}(2013)}]{Burke13}
{Burke}, C., \& {Collins}, C.~A. 2013, \mnras, 434, 2856

\bibitem[{{Butcher} \& {Oemler}(1978)}]{Butcher78}
{Butcher}, H., \& {Oemler}, Jr., A. 1978, \apj, 219, 18

\bibitem[{{Calzetti} {et~al.}(2000){Calzetti}, {Armus}, {Bohlin}, {Kinney},
  {Koornneef}, \& {Storchi-Bergmann}}]{Calzetti00}
{Calzetti}, D., {Armus}, L., {Bohlin}, R.~C., {et~al.} 2000, \apj, 533, 682

\bibitem[{{Capak} {et~al.}(2007{\natexlab{a}}){Capak}, {Abraham}, {Ellis},
  {Mobasher}, {Scoville}, {Sheth}, \& {Koekemoer}}]{Capak07b}
{Capak}, P., {Abraham}, R.~G., {Ellis}, R.~S., {et~al.} 2007{\natexlab{a}},
  \apjs, 172, 284

\bibitem[{{Capak} {et~al.}(2007{\natexlab{b}}){Capak}, {Aussel}, {Ajiki},
  {McCracken}, {Mobasher}, {Scoville}, {Shopbell}, {Taniguchi}, {Thompson},
  {Tribiano}, {Sasaki}, {Blain}, {Brusa}, {Carilli}, {Comastri}, {Carollo},
  {Cassata}, {Colbert}, {Ellis}, {Elvis}, {Giavalisco}, {Green}, {Guzzo},
  {Hasinger}, {Ilbert}, {Impey}, {Jahnke}, {Kartaltepe}, {Kneib}, {Koda},
  {Koekemoer}, {Komiyama}, {Leauthaud}, {Le Fevre}, {Lilly}, {Liu}, {Massey},
  {Miyazaki}, {Murayama}, {Nagao}, {Peacock}, {Pickles}, {Porciani}, {Renzini},
  {Rhodes}, {Rich}, {Salvato}, {Sanders}, {Scarlata}, {Schiminovich},
  {Schinnerer}, {Scodeggio}, {Sheth}, {Shioya}, {Tasca}, {Taylor}, {Yan}, \&
  {Zamorani}}]{Capak07}
{Capak}, P., {Aussel}, H., {Ajiki}, M., {et~al.} 2007{\natexlab{b}}, \apjs,
  172, 99

\bibitem[{{Carollo} {et~al.}(2016){Carollo}, {Cibinel}, {Lilly}, {Pipino},
  {Bonoli}, {Finoguenov}, {Miniati}, {Norberg}, \& {Silverman}}]{Carollo16}
{Carollo}, C.~M., {Cibinel}, A., {Lilly}, S.~J., {et~al.} 2016, \apj, 818, 180

\bibitem[{{Cen}(2014)}]{Cen14}
{Cen}, R. 2014, \apj, 781, 38

\bibitem[{{Chabrier}(2003)}]{Chabrier03}
{Chabrier}, G. 2003, \pasp, 115, 763

\bibitem[{{Chiang} {et~al.}(2015){Chiang}, {Overzier}, {Gebhardt},
  {Finkelstein}, {Chiang}, {Hill}, {Blanc}, {Drory}, {Chonis}, {Zeimann},
  {Hagen}, {Schneider}, {Jogee}, {Ciardullo}, \& {Gronwall}}]{Chiang15}
{Chiang}, Y.-K., {Overzier}, R.~A., {Gebhardt}, K., {et~al.} 2015, \apj, 808,
  37

\bibitem[{{Cooper} {et~al.}(2005){Cooper}, {Newman}, {Madgwick}, {Gerke},
  {Yan}, \& {Davis}}]{Cooper05}
{Cooper}, M.~C., {Newman}, J.~A., {Madgwick}, D.~S., {et~al.} 2005, \apj, 634,
  833

\bibitem[{{Cooper} {et~al.}(2008){Cooper}, {Newman}, {Weiner}, {Yan},
  {Willmer}, {Bundy}, {Coil}, {Conselice}, {Davis}, {Faber}, {Gerke},
  {Guhathakurta}, {Koo}, \& {Noeske}}]{Cooper08}
{Cooper}, M.~C., {Newman}, J.~A., {Weiner}, B.~J., {et~al.} 2008, \mnras, 383,
  1058

\bibitem[{{Cooper} {et~al.}(2012){Cooper}, {Griffith}, {Newman}, {Coil},
  {Davis}, {Dutton}, {Faber}, {Guhathakurta}, {Koo}, {Lotz}, {Weiner},
  {Willmer}, \& {Yan}}]{Cooper12a}
{Cooper}, M.~C., {Griffith}, R.~L., {Newman}, J.~A., {et~al.} 2012, \mnras,
  419, 3018

\bibitem[{{Couch} \& {Sharples}(1987)}]{Couch87}
{Couch}, W.~J., \& {Sharples}, R.~M. 1987, \mnras, 229, 423

\bibitem[{{Cowie} \& {Songaila}(1977)}]{Cowie77}
{Cowie}, L.~L., \& {Songaila}, A. 1977, \nat, 266, 501

\bibitem[{{Cowie} {et~al.}(1996){Cowie}, {Songaila}, {Hu}, \&
  {Cohen}}]{Cowie96}
{Cowie}, L.~L., {Songaila}, A., {Hu}, E.~M., \& {Cohen}, J.~G. 1996, \aj, 112,
  839

\bibitem[{{Daddi} {et~al.}(2007){Daddi}, {Dickinson}, {Morrison}, {Chary},
  {Cimatti}, {Elbaz}, {Frayer}, {Renzini}, {Pope}, {Alexander}, {Bauer},
  {Giavalisco}, {Huynh}, {Kurk}, \& {Mignoli}}]{Daddi07}
{Daddi}, E., {Dickinson}, M., {Morrison}, G., {et~al.} 2007, \apj, 670, 156

\bibitem[{{Darvish} {et~al.}(2015{\natexlab{a}}){Darvish}, {Mobasher},
  {Sobral}, {Hemmati}, {Nayyeri}, \& {Shivaei}}]{Darvish15b}
{Darvish}, B., {Mobasher}, B., {Sobral}, D., {et~al.} 2015{\natexlab{a}}, \apj,
  814, 84

\bibitem[{{Darvish} {et~al.}(2015{\natexlab{b}}){Darvish}, {Mobasher},
  {Sobral}, {Scoville}, \& {Aragon-Calvo}}]{Darvish15a}
{Darvish}, B., {Mobasher}, B., {Sobral}, D., {Scoville}, N., \& {Aragon-Calvo},
  M. 2015{\natexlab{b}}, \apj, 805, 121

\bibitem[{{Darvish} {et~al.}(2014){Darvish}, {Sobral}, {Mobasher}, {Scoville},
  {Best}, {Sales}, \& {Smail}}]{Darvish14}
{Darvish}, B., {Sobral}, D., {Mobasher}, B., {et~al.} 2014, \apj, 796, 51

\bibitem[{{Davidzon} {et~al.}(2016){Davidzon}, {Cucciati}, {Bolzonella}, {De
  Lucia}, {Zamorani}, {Arnouts}, {Moutard}, {Ilbert}, {Garilli}, {Scodeggio},
  {Guzzo}, {Abbas}, {Adami}, {Bel}, {Bottini}, {Branchini}, {Cappi}, {Coupon},
  {de la Torre}, {Di Porto}, {Fritz}, {Franzetti}, {Fumana}, {Granett},
  {Guennou}, {Iovino}, {Krywult}, {Le Brun}, {Le F{\`e}vre}, {Maccagni},
  {Ma{\l}ek}, {Marulli}, {McCracken}, {Mellier}, {Moscardini}, {Polletta},
  {Pollo}, {Tasca}, {Tojeiro}, {Vergani}, \& {Zanichelli}}]{Davidzon16}
{Davidzon}, I., {Cucciati}, O., {Bolzonella}, M., {et~al.} 2016, \aap, 586, A23

\bibitem[{{De Lucia} {et~al.}(2012){De Lucia}, {Weinmann}, {Poggianti},
  {Arag{\'o}n-Salamanca}, \& {Zaritsky}}]{DeLucia12}
{De Lucia}, G., {Weinmann}, S., {Poggianti}, B.~M., {Arag{\'o}n-Salamanca}, A.,
  \& {Zaritsky}, D. 2012, \mnras, 423, 1277

\bibitem[{{De Lucia} {et~al.}(2007){De Lucia}, {Poggianti},
  {Arag{\'o}n-Salamanca}, {White}, {Zaritsky}, {Clowe}, {Halliday}, {Jablonka},
  {von der Linden}, {Milvang-Jensen}, {Pell{\'o}}, {Rudnick}, {Saglia}, \&
  {Simard}}]{Delucia07}
{De Lucia}, G., {Poggianti}, B.~M., {Arag{\'o}n-Salamanca}, A., {et~al.} 2007,
  \mnras, 374, 809

\bibitem[{{Dekel} \& {Birnboim}(2006)}]{Dekel06}
{Dekel}, A., \& {Birnboim}, Y. 2006, \mnras, 368, 2

\bibitem[{{Delaye} {et~al.}(2014){Delaye}, {Huertas-Company}, {Mei}, {Lidman},
  {Licitra}, {Newman}, {Raichoor}, {Shankar}, {Barrientos}, {Bernardi},
  {Cerulo}, {Couch}, {Demarco}, {Mu{\~n}oz}, {S{\'a}nchez-Janssen}, \&
  {Tanaka}}]{Delaye14}
{Delaye}, L., {Huertas-Company}, M., {Mei}, S., {et~al.} 2014, \mnras, 441, 203

\bibitem[{{Donoso} {et~al.}(2014){Donoso}, {Yan}, {Stern}, \&
  {Assef}}]{Donoso14}
{Donoso}, E., {Yan}, L., {Stern}, D., \& {Assef}, R.~J. 2014, \apj, 789, 44

\bibitem[{{Dressler} \& {Gunn}(1983)}]{Dressler83}
{Dressler}, A., \& {Gunn}, J.~E. 1983, \apj, 270, 7

\bibitem[{{Dressler} {et~al.}(1994){Dressler}, {Oemler}, {Butcher}, \&
  {Gunn}}]{Dressler94}
{Dressler}, A., {Oemler}, Jr., A., {Butcher}, H.~R., \& {Gunn}, J.~E. 1994,
  \apj, 430, 107

\bibitem[{{Dressler} {et~al.}(1999){Dressler}, {Smail}, {Poggianti}, {Butcher},
  {Couch}, {Ellis}, \& {Oemler}}]{Dressler99}
{Dressler}, A., {Smail}, I., {Poggianti}, B.~M., {et~al.} 1999, \apjs, 122, 51

\bibitem[{{Elbaz} {et~al.}(2007){Elbaz}, {Daddi}, {Le Borgne}, {Dickinson},
  {Alexander}, {Chary}, {Starck}, {Brandt}, {Kitzbichler}, {MacDonald},
  {Nonino}, {Popesso}, {Stern}, \& {Vanzella}}]{Elbaz07}
{Elbaz}, D., {Daddi}, E., {Le Borgne}, D., {et~al.} 2007, \aap, 468, 33

\bibitem[{{Erfanianfar} {et~al.}(2016){Erfanianfar}, {Popesso}, {Finoguenov},
  {Wilman}, {Wuyts}, {Biviano}, {Salvato}, {Mirkazemi}, {Morselli}, {Ziparo},
  {Nandra}, {Lutz}, {Elbaz}, {Dickinson}, {Tanaka}, {Altieri}, {Aussel},
  {Bauer}, {Berta}, {Bielby}, {Brandt}, {Cappelluti}, {Cimatti}, {Cooper},
  {Fadda}, {Ilbert}, {Le Floch}, {Magnelli}, {Mulchaey}, {Nordon}, {Newman},
  {Poglitsch}, \& {Pozzi}}]{Erfanianfar16}
{Erfanianfar}, G., {Popesso}, P., {Finoguenov}, A., {et~al.} 2016, \mnras, 455,
  2839

\bibitem[{{Fabian}(2012)}]{Fabian12}
{Fabian}, A.~C. 2012, \araa, 50, 455

\bibitem[{{Fakhouri} \& {Ma}(2009)}]{Fakhouri09}
{Fakhouri}, O., \& {Ma}, C.-P. 2009, \mnras, 394, 1825

\bibitem[{{Farouki} \& {Shapiro}(1981)}]{Farouki81}
{Farouki}, R., \& {Shapiro}, S.~L. 1981, \apj, 243, 32

\bibitem[{{Foltz} {et~al.}(2015){Foltz}, {Rettura}, {Wilson}, {van der Burg},
  {Muzzin}, {Lidman}, {Demarco}, {Nantais}, {DeGroot}, \& {Yee}}]{Foltz15}
{Foltz}, R., {Rettura}, A., {Wilson}, G., {et~al.} 2015, \apj, 812, 138

\bibitem[{{Fontana} {et~al.}(2009){Fontana}, {Santini}, {Grazian},
  {Pentericci}, {Fiore}, {Castellano}, {Giallongo}, {Menci}, {Salimbeni},
  {Cristiani}, {Nonino}, \& {Vanzella}}]{Fontana09}
{Fontana}, A., {Santini}, P., {Grazian}, A., {et~al.} 2009, \aap, 501, 15

\bibitem[{{Gabor} \& {Dav{\'e}}(2015)}]{Gabor15}
{Gabor}, J.~M., \& {Dav{\'e}}, R. 2015, \mnras, 447, 374

\bibitem[{{Genel}(2016)}]{Genel16}
{Genel}, S. 2016, ArXiv e-prints, arXiv:1602.02773

\bibitem[{{Gr{\"u}tzbauch} {et~al.}(2011){Gr{\"u}tzbauch}, {Conselice},
  {Bauer}, {Bluck}, {Chuter}, {Buitrago}, {Mortlock}, {Weinzirl}, \&
  {Jogee}}]{Grutzbauch11}
{Gr{\"u}tzbauch}, R., {Conselice}, C.~J., {Bauer}, A.~E., {et~al.} 2011,
  \mnras, 418, 938

\bibitem[{{Gunn} \& {Gott}(1972)}]{Gunn72}
{Gunn}, J.~E., \& {Gott}, III, J.~R. 1972, \apj, 176, 1

\bibitem[{{Haines} {et~al.}(2013){Haines}, {Pereira}, {Smith}, {Egami},
  {Sanderson}, {Babul}, {Finoguenov}, {Merluzzi}, {Busarello}, {Rawle}, \&
  {Okabe}}]{Haines13}
{Haines}, C.~P., {Pereira}, M.~J., {Smith}, G.~P., {et~al.} 2013, \apj, 775,
  126

\bibitem[{{Hartley} {et~al.}(2015){Hartley}, {Conselice}, {Mortlock},
  {Foucaud}, \& {Simpson}}]{Hartley15}
{Hartley}, W.~G., {Conselice}, C.~J., {Mortlock}, A., {Foucaud}, S., \&
  {Simpson}, C. 2015, \mnras, 451, 1613

\bibitem[{{Hayashi} {et~al.}(2014){Hayashi}, {Kodama}, {Koyama}, {Tadaki},
  {Tanaka}, {Shimakawa}, {Matsuda}, {Sobral}, {Best}, \& {Smail}}]{Hayashi14}
{Hayashi}, M., {Kodama}, T., {Koyama}, Y., {et~al.} 2014, \mnras, 439, 2571

\bibitem[{{Hayward} \& {Hopkins}(2015)}]{Hayward15}
{Hayward}, C.~C., \& {Hopkins}, P.~F. 2015, ArXiv e-prints, arXiv:1510.05650

\bibitem[{{Hirschmann} {et~al.}(2014){Hirschmann}, {De Lucia}, {Wilman},
  {Weinmann}, {Iovino}, {Cucciati}, {Zibetti}, \& {Villalobos}}]{Hirschmann14}
{Hirschmann}, M., {De Lucia}, G., {Wilman}, D., {et~al.} 2014, \mnras, 444,
  2938

\bibitem[{{Hopkins} {et~al.}(2014){Hopkins}, {Kere{\v s}}, {O{\~n}orbe},
  {Faucher-Gigu{\`e}re}, {Quataert}, {Murray}, \& {Bullock}}]{Hopkins14}
{Hopkins}, P.~F., {Kere{\v s}}, D., {O{\~n}orbe}, J., {et~al.} 2014, \mnras,
  445, 581

\bibitem[{{Ilbert} {et~al.}(2009){Ilbert}, {Capak}, {Salvato}, {Aussel},
  {McCracken}, {Sanders}, {Scoville}, {Kartaltepe}, {Arnouts}, {Le Floc'h},
  {Mobasher}, {Taniguchi}, {Lamareille}, {Leauthaud}, {Sasaki}, {Thompson},
  {Zamojski}, {Zamorani}, {Bardelli}, {Bolzonella}, {Bongiorno}, {Brusa},
  {Caputi}, {Carollo}, {Contini}, {Cook}, {Coppa}, {Cucciati}, {de la Torre},
  {de Ravel}, {Franzetti}, {Garilli}, {Hasinger}, {Iovino}, {Kampczyk},
  {Kneib}, {Knobel}, {Kovac}, {Le Borgne}, {Le Brun}, {F{\`e}vre}, {Lilly},
  {Looper}, {Maier}, {Mainieri}, {Mellier}, {Mignoli}, {Murayama}, {Pell{\`o}},
  {Peng}, {P{\'e}rez-Montero}, {Renzini}, {Ricciardelli}, {Schiminovich},
  {Scodeggio}, {Shioya}, {Silverman}, {Surace}, {Tanaka}, {Tasca}, {Tresse},
  {Vergani}, \& {Zucca}}]{Ilbert09}
{Ilbert}, O., {Capak}, P., {Salvato}, M., {et~al.} 2009, \apj, 690, 1236

\bibitem[{{Ilbert} {et~al.}(2013){Ilbert}, {McCracken}, {Le F{\`e}vre},
  {Capak}, {Dunlop}, {Karim}, {Renzini}, {Caputi}, {Boissier}, {Arnouts},
  {Aussel}, {Comparat}, {Guo}, {Hudelot}, {Kartaltepe}, {Kneib}, {Krogager},
  {Le Floc'h}, {Lilly}, {Mellier}, {Milvang-Jensen}, {Moutard}, {Onodera},
  {Richard}, {Salvato}, {Sanders}, {Scoville}, {Silverman}, {Taniguchi},
  {Tasca}, {Thomas}, {Toft}, {Tresse}, {Vergani}, {Wolk}, \& {Zirm}}]{Ilbert13}
{Ilbert}, O., {McCracken}, H.~J., {Le F{\`e}vre}, O., {et~al.} 2013, \aap, 556,
  A55

\bibitem[{{Karim} {et~al.}(2011){Karim}, {Schinnerer},
  {Mart{\'{\i}}nez-Sansigre}, {Sargent}, {van der Wel}, {Rix}, {Ilbert},
  {Smol{\v c}i{\'c}}, {Carilli}, {Pannella}, {Koekemoer}, {Bell}, \&
  {Salvato}}]{Karim11}
{Karim}, A., {Schinnerer}, E., {Mart{\'{\i}}nez-Sansigre}, A., {et~al.} 2011,
  \apj, 730, 61

\bibitem[{{Kauffmann} {et~al.}(2004){Kauffmann}, {White}, {Heckman},
  {M{\'e}nard}, {Brinchmann}, {Charlot}, {Tremonti}, \&
  {Brinkmann}}]{Kauffmann04}
{Kauffmann}, G., {White}, S.~D.~M., {Heckman}, T.~M., {et~al.} 2004, \mnras,
  353, 713

\bibitem[{{Kawinwanichakij} {et~al.}(2014){Kawinwanichakij}, {Papovich},
  {Quadri}, {Tran}, {Spitler}, {Kacprzak}, {Labb{\'e}}, {Straatman},
  {Glazebrook}, {Allen}, {Cowley}, {Dav{\'e}}, {Dekel}, {Ferguson}, {Hartley},
  {Koekemoer}, {Koo}, {Lu}, {Mehrtens}, {Nanayakkara}, {Persson}, {Rees},
  {Salmon}, {Tilvi}, {Tomczak}, \& {van Dokkum}}]{Kawinwanichakij14}
{Kawinwanichakij}, L., {Papovich}, C., {Quadri}, R.~F., {et~al.} 2014, \apj,
  792, 103

\bibitem[{{Kennicutt}(1998)}]{Kennicutt98}
{Kennicutt}, Jr., R.~C. 1998, \araa, 36, 189

\bibitem[{{Kewley} {et~al.}(2006){Kewley}, {Geller}, \& {Barton}}]{Kewley06}
{Kewley}, L.~J., {Geller}, M.~J., \& {Barton}, E.~J. 2006, \aj, 131, 2004

\bibitem[{{Khostovan} {et~al.}(2015){Khostovan}, {Sobral}, {Mobasher}, {Best},
  {Smail}, {Stott}, {Hemmati}, \& {Nayyeri}}]{Khostovan15}
{Khostovan}, A.~A., {Sobral}, D., {Mobasher}, B., {et~al.} 2015, \mnras, 452,
  3948

\bibitem[{{Knobel} {et~al.}(2015){Knobel}, {Lilly}, {Woo}, \& {Kova{\v
  c}}}]{Knobel15}
{Knobel}, C., {Lilly}, S.~J., {Woo}, J., \& {Kova{\v c}}, K. 2015, \apj, 800,
  24

\bibitem[{{Knobel} {et~al.}(2013){Knobel}, {Lilly}, {Kova{\v c}}, {Peng},
  {Bschorr}, {Carollo}, {Contini}, {Kneib}, {Le Fevre}, {Mainieri}, {Renzini},
  {Scodeggio}, {Zamorani}, {Bardelli}, {Bolzonella}, {Bongiorno}, {Caputi},
  {Cucciati}, {de la Torre}, {de Ravel}, {Franzetti}, {Garilli}, {Iovino},
  {Kampczyk}, {Lamareille}, {Le Borgne}, {Le Brun}, {Maier}, {Mignoli},
  {Pello}, {Perez Montero}, {Presotto}, {Silverman}, {Tanaka}, {Tasca},
  {Tresse}, {Vergani}, {Zucca}, {Barnes}, {Bordoloi}, {Cappi}, {Cimatti},
  {Coppa}, {Koekemoer}, {L{\'o}pez-Sanjuan}, {McCracken}, {Moresco}, {Nair},
  {Pozzetti}, \& {Welikala}}]{Knobel13}
{Knobel}, C., {Lilly}, S.~J., {Kova{\v c}}, K., {et~al.} 2013, \apj, 769, 24

\bibitem[{{Koyama} {et~al.}(2014){Koyama}, {Kodama}, {Tadaki}, {Hayashi},
  {Tanaka}, \& {Shimakawa}}]{Koyama14}
{Koyama}, Y., {Kodama}, T., {Tadaki}, K.-i., {et~al.} 2014, \apj, 789, 18

\bibitem[{{Koyama} {et~al.}(2013){Koyama}, {Smail}, {Kurk}, {Geach}, {Sobral},
  {Kodama}, {Nakata}, {Swinbank}, {Best}, {Hayashi}, \& {Tadaki}}]{Koyama13a}
{Koyama}, Y., {Smail}, I., {Kurk}, J., {et~al.} 2013, \mnras, 434, 423

\bibitem[{{Laigle} {et~al.}(2016){Laigle}, {McCracken}, {Ilbert}, {Hsieh},
  {Davidzon}, {Capak}, {Hasinger}, {Silverman}, {Pichon}, {Coupon}, {Aussel},
  {Le Borgne}, {Caputi}, {Cassata}, {Chang}, {Civano}, {Dunlop}, {Fynbo},
  {kartaltepe}, {Koekemoer}, {Le Fevre}, {Le Floc'h}, {Leauthaud}, {Lilly},
  {Lin}, {Marchesi}, {Milvang-Jensen}, {Salvato}, {Sanders}, {Scoville},
  {Smolcic}, {Stockmann}, {Taniguchi}, {Tasca}, {Toft}, {Vaccari}, \&
  {Zabl}}]{Laigle16}
{Laigle}, C., {McCracken}, H.~J., {Ilbert}, O., {et~al.} 2016, ArXiv e-prints,
  arXiv:1604.02350

\bibitem[{{Lani} {et~al.}(2013){Lani}, {Almaini}, {Hartley}, {Mortlock},
  {H{\"a}u{\ss}ler}, {Chuter}, {Simpson}, {van der Wel}, {Gr{\"u}tzbauch},
  {Conselice}, {Bradshaw}, {Cooper}, {Faber}, {Grogin}, {Kocevski},
  {Koekemoer}, \& {Lai}}]{Lani13}
{Lani}, C., {Almaini}, O., {Hartley}, W.~G., {et~al.} 2013, \mnras, 435, 207

\bibitem[{{Larson} {et~al.}(1980){Larson}, {Tinsley}, \& {Caldwell}}]{Larson80}
{Larson}, R.~B., {Tinsley}, B.~M., \& {Caldwell}, C.~N. 1980, \apj, 237, 692

\bibitem[{{Lee} {et~al.}(2015){Lee}, {Im}, {Kim}, {Lotz}, {McPartland}, {Peth},
  \& {Koekemoer}}]{Lee15}
{Lee}, S.-K., {Im}, M., {Kim}, J.-W., {et~al.} 2015, \apj, 810, 90

\bibitem[{{Lidman} {et~al.}(2012){Lidman}, {Suherli}, {Muzzin}, {Wilson},
  {Demarco}, {Brough}, {Rettura}, {Cox}, {DeGroot}, {Yee}, {Gilbank},
  {Hoekstra}, {Balogh}, {Ellingson}, {Hicks}, {Nantais}, {Noble}, {Lacy},
  {Surace}, \& {Webb}}]{Lidman12}
{Lidman}, C., {Suherli}, J., {Muzzin}, A., {et~al.} 2012, \mnras, 427, 550

\bibitem[{{Lin} {et~al.}(2014){Lin}, {Jian}, {Foucaud}, {Norberg}, {Bower},
  {Cole}, {Arnalte-Mur}, {Chen}, {Coupon}, {Hsieh}, {Heinis}, {Phleps}, {Chen},
  {Lee}, {Burgett}, {Chambers}, {Denneau}, {Draper}, {Flewelling}, {Hodapp},
  {Huber}, {Kaiser}, {Kudritzki}, {Magnier}, {Metcalfe}, {Price}, {Tonry},
  {Wainscoat}, \& {Waters}}]{Lin14}
{Lin}, L., {Jian}, H.-Y., {Foucaud}, S., {et~al.} 2014, \apj, 782, 33

\bibitem[{{Lin} {et~al.}(2013){Lin}, {Brodwin}, {Gonzalez}, {Bode},
  {Eisenhardt}, {Stanford}, \& {Vikhlinin}}]{Lin13}
{Lin}, Y.-T., {Brodwin}, M., {Gonzalez}, A.~H., {et~al.} 2013, \apj, 771, 61

\bibitem[{{Madau} \& {Dickinson}(2014)}]{Madau14}
{Madau}, P., \& {Dickinson}, M. 2014, \araa, 52, 415

\bibitem[{{Malavasi} {et~al.}(2016){Malavasi}, {Pozzetti}, {Cucciati},
  {Bardelli}, \& {Cimatti}}]{Malavasi16}
{Malavasi}, N., {Pozzetti}, L., {Cucciati}, O., {Bardelli}, S., \& {Cimatti},
  A. 2016, \aap, 585, A116

\bibitem[{{Martini} {et~al.}(2006){Martini}, {Kelson}, {Kim}, {Mulchaey}, \&
  {Athey}}]{Martini06}
{Martini}, P., {Kelson}, D.~D., {Kim}, E., {Mulchaey}, J.~S., \& {Athey}, A.~A.
  2006, \apj, 644, 116

\bibitem[{{Martini} {et~al.}(2013){Martini}, {Miller}, {Brodwin}, {Stanford},
  {Gonzalez}, {Bautz}, {Hickox}, {Stern}, {Eisenhardt}, {Galametz}, {Norman},
  {Jannuzi}, {Dey}, {Murray}, {Jones}, \& {Brown}}]{Martini13}
{Martini}, P., {Miller}, E.~D., {Brodwin}, M., {et~al.} 2013, \apj, 768, 1

\bibitem[{{McCracken} {et~al.}(2012){McCracken}, {Milvang-Jensen}, {Dunlop},
  {Franx}, {Fynbo}, {Le F{\`e}vre}, {Holt}, {Caputi}, {Goranova}, {Buitrago},
  {Emerson}, {Freudling}, {Hudelot}, {L{\'o}pez-Sanjuan}, {Magnard}, {Mellier},
  {M{\o}ller}, {Nilsson}, {Sutherland}, {Tasca}, \& {Zabl}}]{McCracken12}
{McCracken}, H.~J., {Milvang-Jensen}, B., {Dunlop}, J., {et~al.} 2012, \aap,
  544, A156

\bibitem[{{Merritt}(1983)}]{Merritt83}
{Merritt}, D. 1983, \apj, 264, 24

\bibitem[{{Merritt}(1984)}]{Merritt84}
---. 1984, \apj, 276, 26

\bibitem[{{Mihos} \& {Hernquist}(1996)}]{Mihos96}
{Mihos}, J.~C., \& {Hernquist}, L. 1996, \apj, 464, 641

\bibitem[{{Miller} {et~al.}(2003){Miller}, {Nichol}, {G{\'o}mez}, {Hopkins}, \&
  {Bernardi}}]{Miller03}
{Miller}, C.~J., {Nichol}, R.~C., {G{\'o}mez}, P.~L., {Hopkins}, A.~M., \&
  {Bernardi}, M. 2003, \apj, 597, 142

\bibitem[{{Moore} {et~al.}(1998){Moore}, {Lake}, \& {Katz}}]{Moore98}
{Moore}, B., {Lake}, G., \& {Katz}, N. 1998, \apj, 495, 139

\bibitem[{{Mortlock} {et~al.}(2015){Mortlock}, {Conselice}, {Hartley},
  {Duncan}, {Lani}, {Ownsworth}, {Almaini}, {Wel}, {Huang}, {Ashby}, {Willner},
  {Fontana}, {Dekel}, {Koekemoer}, {Ferguson}, {Faber}, {Grogin}, \&
  {Kocevski}}]{Mortlock15}
{Mortlock}, A., {Conselice}, C.~J., {Hartley}, W.~G., {et~al.} 2015, \mnras,
  447, 2

\bibitem[{{Moster} {et~al.}(2011){Moster}, {Somerville}, {Newman}, \&
  {Rix}}]{Moster11}
{Moster}, B.~P., {Somerville}, R.~S., {Newman}, J.~A., \& {Rix}, H.-W. 2011,
  \apj, 731, 113

\bibitem[{{Mouhcine} {et~al.}(2007){Mouhcine}, {Baldry}, \&
  {Bamford}}]{Mouhcine07}
{Mouhcine}, M., {Baldry}, I.~K., \& {Bamford}, S.~P. 2007, \mnras, 382, 801

\bibitem[{{Muldrew} {et~al.}(2012){Muldrew}, {Croton}, {Skibba}, {Pearce},
  {Ann}, {Baldry}, {Brough}, {Choi}, {Conselice}, {Cowan}, {Gallazzi}, {Gray},
  {Gr{\"u}tzbauch}, {Li}, {Park}, {Pilipenko}, {Podgorzec}, {Robotham},
  {Wilman}, {Yang}, {Zhang}, \& {Zibetti}}]{Muldrew12}
{Muldrew}, S.~I., {Croton}, D.~J., {Skibba}, R.~A., {et~al.} 2012, \mnras, 419,
  2670

\bibitem[{{Muzzin} {et~al.}(2012){Muzzin}, {Wilson}, {Yee}, {Gilbank},
  {Hoekstra}, {Demarco}, {Balogh}, {van Dokkum}, {Franx}, {Ellingson}, {Hicks},
  {Nantais}, {Noble}, {Lacy}, {Lidman}, {Rettura}, {Surace}, \&
  {Webb}}]{Muzzin12}
{Muzzin}, A., {Wilson}, G., {Yee}, H.~K.~C., {et~al.} 2012, \apj, 746, 188

\bibitem[{{Nayyeri} {et~al.}(2014){Nayyeri}, {Mobasher}, {Hemmati}, {De
  Barros}, {Ferguson}, {Wiklind}, {Dahlen}, {Dickinson}, {Giavalisco},
  {Fontana}, {Ashby}, {Barro}, {Guo}, {Hathi}, {Kassin}, {Koekemoer},
  {Willner}, {Dunlop}, {Paris}, \& {Targett}}]{Nayyeri14}
{Nayyeri}, H., {Mobasher}, B., {Hemmati}, S., {et~al.} 2014, \apj, 794, 68

\bibitem[{{Newman} {et~al.}(2014){Newman}, {Ellis}, {Andreon}, {Treu},
  {Raichoor}, \& {Trinchieri}}]{Newman14}
{Newman}, A.~B., {Ellis}, R.~S., {Andreon}, S., {et~al.} 2014, \apj, 788, 51

\bibitem[{{Nulsen}(1982)}]{Nulsen82}
{Nulsen}, P.~E.~J. 1982, \mnras, 198, 1007

\bibitem[{{Omand} {et~al.}(2014){Omand}, {Balogh}, \& {Poggianti}}]{Omand14}
{Omand}, C.~M.~B., {Balogh}, M.~L., \& {Poggianti}, B.~M. 2014, \mnras, 440,
  843

\bibitem[{{Patel} {et~al.}(2009){Patel}, {Holden}, {Kelson}, {Illingworth}, \&
  {Franx}}]{Patel09}
{Patel}, S.~G., {Holden}, B.~P., {Kelson}, D.~D., {Illingworth}, G.~D., \&
  {Franx}, M. 2009, \apjl, 705, L67

\bibitem[{{Patel} {et~al.}(2011){Patel}, {Kelson}, {Holden}, {Franx}, \&
  {Illingworth}}]{Patel11}
{Patel}, S.~G., {Kelson}, D.~D., {Holden}, B.~P., {Franx}, M., \&
  {Illingworth}, G.~D. 2011, \apj, 735, 53

\bibitem[{{Patton} \& {Atfield}(2008)}]{Patton08}
{Patton}, D.~R., \& {Atfield}, J.~E. 2008, \apj, 685, 235

\bibitem[{{Peng} {et~al.}(2012){Peng}, {Lilly}, {Renzini}, \&
  {Carollo}}]{Peng12}
{Peng}, Y.-j., {Lilly}, S.~J., {Renzini}, A., \& {Carollo}, M. 2012, \apj, 757,
  4

\bibitem[{{Peng} {et~al.}(2010){Peng}, {Lilly}, {Kova{\v c}}, {Bolzonella},
  {Pozzetti}, {Renzini}, {Zamorani}, {Ilbert}, {Knobel}, {Iovino}, {Maier},
  {Cucciati}, {Tasca}, {Carollo}, {Silverman}, {Kampczyk}, {de Ravel},
  {Sanders}, {Scoville}, {Contini}, {Mainieri}, {Scodeggio}, {Kneib}, {Le
  F{\`e}vre}, {Bardelli}, {Bongiorno}, {Caputi}, {Coppa}, {de la Torre},
  {Franzetti}, {Garilli}, {Lamareille}, {Le Borgne}, {Le Brun}, {Mignoli},
  {Perez Montero}, {Pello}, {Ricciardelli}, {Tanaka}, {Tresse}, {Vergani},
  {Welikala}, {Zucca}, {Oesch}, {Abbas}, {Barnes}, {Bordoloi}, {Bottini},
  {Cappi}, {Cassata}, {Cimatti}, {Fumana}, {Hasinger}, {Koekemoer},
  {Leauthaud}, {Maccagni}, {Marinoni}, {McCracken}, {Memeo}, {Meneux}, {Nair},
  {Porciani}, {Presotto}, \& {Scaramella}}]{Peng10}
{Peng}, Y.-j., {Lilly}, S.~J., {Kova{\v c}}, K., {et~al.} 2010, \apj, 721, 193

\bibitem[{{Poggianti} {et~al.}(2009){Poggianti}, {Arag{\'o}n-Salamanca},
  {Zaritsky}, {De Lucia}, {Milvang-Jensen}, {Desai}, {Jablonka}, {Halliday},
  {Rudnick}, {Varela}, {Bamford}, {Best}, {Clowe}, {Noll}, {Saglia},
  {Pell{\'o}}, {Simard}, {von der Linden}, \& {White}}]{Poggianti09}
{Poggianti}, B.~M., {Arag{\'o}n-Salamanca}, A., {Zaritsky}, D., {et~al.} 2009,
  \apj, 693, 112

\bibitem[{{Popesso} \& {Biviano}(2006)}]{Popesso06}
{Popesso}, P., \& {Biviano}, A. 2006, \aap, 460, L23

\bibitem[{{Popesso} {et~al.}(2011){Popesso}, {Rodighiero}, {Saintonge},
  {Santini}, {Grazian}, {Lutz}, {Brusa}, {Altieri}, {Andreani}, {Aussel},
  {Berta}, {Bongiovanni}, {Cava}, {Cepa}, {Cimatti}, {Daddi}, {Dominguez},
  {Elbaz}, {F{\"o}rster Schreiber}, {Genzel}, {Gruppioni}, {Magdis},
  {Maiolino}, {Magnelli}, {Nordon}, {P{\'e}rez Garc{\'{\i}}a}, {Poglitsch},
  {Pozzi}, {Riguccini}, {Sanchez-Portal}, {Shao}, {Sturm}, {Tacconi},
  {Valtchanov}, {Wieprecht}, \& {Wetzstein}}]{Popesso11}
{Popesso}, P., {Rodighiero}, G., {Saintonge}, A., {et~al.} 2011, \aap, 532,
  A145

\bibitem[{{Postman} {et~al.}(2005){Postman}, {Franx}, {Cross}, {Holden},
  {Ford}, {Illingworth}, {Goto}, {Demarco}, {Rosati}, {Blakeslee}, {Tran},
  {Ben{\'{\i}}tez}, {Clampin}, {Hartig}, {Homeier}, {Ardila}, {Bartko},
  {Bouwens}, {Bradley}, {Broadhurst}, {Brown}, {Burrows}, {Cheng}, {Feldman},
  {Golimowski}, {Gronwall}, {Infante}, {Kimble}, {Krist}, {Lesser}, {Martel},
  {Mei}, {Menanteau}, {Meurer}, {Miley}, {Motta}, {Sirianni}, {Sparks}, {Tran},
  {Tsvetanov}, {White}, \& {Zheng}}]{Postman05}
{Postman}, M., {Franx}, M., {Cross}, N.~J.~G., {et~al.} 2005, \apj, 623, 721

\bibitem[{{Pozzetti} {et~al.}(2010){Pozzetti}, {Bolzonella}, {Zucca},
  {Zamorani}, {Lilly}, {Renzini}, {Moresco}, {Mignoli}, {Cassata}, {Tasca},
  {Lamareille}, {Maier}, {Meneux}, {Halliday}, {Oesch}, {Vergani}, {Caputi},
  {Kova{\v c}}, {Cimatti}, {Cucciati}, {Iovino}, {Peng}, {Carollo}, {Contini},
  {Kneib}, {Le F{\'e}vre}, {Mainieri}, {Scodeggio}, {Bardelli}, {Bongiorno},
  {Coppa}, {de la Torre}, {de Ravel}, {Franzetti}, {Garilli}, {Kampczyk},
  {Knobel}, {Le Borgne}, {Le Brun}, {Pell{\`o}}, {Perez Montero},
  {Ricciardelli}, {Silverman}, {Tanaka}, {Tresse}, {Abbas}, {Bottini}, {Cappi},
  {Guzzo}, {Koekemoer}, {Leauthaud}, {Maccagni}, {Marinoni}, {McCracken},
  {Memeo}, {Porciani}, {Scaramella}, {Scarlata}, \& {Scoville}}]{Pozzetti10}
{Pozzetti}, L., {Bolzonella}, M., {Zucca}, E., {et~al.} 2010, \aap, 523, A13

\bibitem[{{Quadri} {et~al.}(2012){Quadri}, {Williams}, {Franx}, \&
  {Hildebrandt}}]{Quadri12}
{Quadri}, R.~F., {Williams}, R.~J., {Franx}, M., \& {Hildebrandt}, H. 2012,
  \apj, 744, 88

\bibitem[{{Rettura} {et~al.}(2010){Rettura}, {Rosati}, {Nonino}, {Fosbury},
  {Gobat}, {Menci}, {Strazzullo}, {Mei}, {Demarco}, \& {Ford}}]{Rettura10}
{Rettura}, A., {Rosati}, P., {Nonino}, M., {et~al.} 2010, \apj, 709, 512

\bibitem[{{Rettura} {et~al.}(2011){Rettura}, {Mei}, {Stanford}, {Raichoor},
  {Moran}, {Holden}, {Rosati}, {Ellis}, {Nakata}, {Nonino}, {Treu},
  {Blakeslee}, {Demarco}, {Eisenhardt}, {Ford}, {Fosbury}, {Illingworth},
  {Huertas-Company}, {Jee}, {Kodama}, {Postman}, {Tanaka}, \&
  {White}}]{Rettura11}
{Rettura}, A., {Mei}, S., {Stanford}, S.~A., {et~al.} 2011, \apj, 732, 94

\bibitem[{{Ricciardelli} {et~al.}(2014){Ricciardelli}, {Cava}, {Varela}, \&
  {Quilis}}]{Ricciardelli14}
{Ricciardelli}, E., {Cava}, A., {Varela}, J., \& {Quilis}, V. 2014, \mnras,
  445, 4045

\bibitem[{{Santos} {et~al.}(2015){Santos}, {Altieri}, {Valtchanov}, {Nastasi},
  {B{\"o}hringer}, {Cresci}, {Elbaz}, {Fassbender}, {Rosati}, {Tozzi}, \&
  {Verdugo}}]{Santos15}
{Santos}, J.~S., {Altieri}, B., {Valtchanov}, I., {et~al.} 2015, \mnras, 447,
  L65

\bibitem[{{Scoville} {et~al.}(2007){Scoville}, {Aussel}, {Brusa}, {Capak},
  {Carollo}, {Elvis}, {Giavalisco}, {Guzzo}, {Hasinger}, {Impey}, {Kneib},
  {LeFevre}, {Lilly}, {Mobasher}, {Renzini}, {Rich}, {Sanders}, {Schinnerer},
  {Schminovich}, {Shopbell}, {Taniguchi}, \& {Tyson}}]{Scoville07}
{Scoville}, N., {Aussel}, H., {Brusa}, M., {et~al.} 2007, \apjs, 172, 1

\bibitem[{{Scoville} {et~al.}(2013){Scoville}, {Arnouts}, {Aussel}, {Benson},
  {Bongiorno}, {Bundy}, {Calvo}, {Capak}, {Carollo}, {Civano}, {Dunlop},
  {Elvis}, {Faisst}, {Finoguenov}, {Fu}, {Giavalisco}, {Guo}, {Ilbert},
  {Iovino}, {Kajisawa}, {Kartaltepe}, {Leauthaud}, {Le F{\`e}vre}, {LeFloch},
  {Lilly}, {Liu}, {Manohar}, {Massey}, {Masters}, {McCracken}, {Mobasher},
  {Peng}, {Renzini}, {Rhodes}, {Salvato}, {Sanders}, {Sarvestani}, {Scarlata},
  {Schinnerer}, {Sheth}, {Shopbell}, {Smol{\v c}i{\'c}}, {Taniguchi}, {Taylor},
  {White}, \& {Yan}}]{Scoville13}
{Scoville}, N., {Arnouts}, S., {Aussel}, H., {et~al.} 2013, \apjs, 206, 3

\bibitem[{{Shankar} {et~al.}(2015){Shankar}, {Buchan}, {Rettura}, {Bouillot},
  {Moreno}, {Licitra}, {Bernardi}, {Huertas-Company}, {Mei}, {Ascaso}, {Sheth},
  {Delaye}, \& {Raichoor}}]{Shankar15}
{Shankar}, F., {Buchan}, S., {Rettura}, A., {et~al.} 2015, \apj, 802, 73

\bibitem[{{Shimakawa} {et~al.}(2015){Shimakawa}, {Kodama}, {Tadaki}, {Hayashi},
  {Koyama}, \& {Tanaka}}]{Shimakawa15}
{Shimakawa}, R., {Kodama}, T., {Tadaki}, K.-i., {et~al.} 2015, \mnras, 448, 666

\bibitem[{{Skelton} {et~al.}(2009){Skelton}, {Bell}, \&
  {Somerville}}]{Skelton09}
{Skelton}, R.~E., {Bell}, E.~F., \& {Somerville}, R.~S. 2009, \apjl, 699, L9

\bibitem[{{Smail} {et~al.}(2014){Smail}, {Geach}, {Swinbank}, {Tadaki},
  {Arumugam}, {Hartley}, {Almaini}, {Bremer}, {Chapin}, {Chapman}, {Danielson},
  {Edge}, {Scott}, {Simpson}, {Simpson}, {Conselice}, {Dunlop}, {Ivison},
  {Karim}, {Kodama}, {Mortlock}, {Robson}, {Roseboom}, {Thomson}, {van der
  Werf}, \& {Webb}}]{Smail14}
{Smail}, I., {Geach}, J.~E., {Swinbank}, A.~M., {et~al.} 2014, \apj, 782, 19

\bibitem[{{Sobral} {et~al.}(2011){Sobral}, {Best}, {Smail}, {Geach},
  {Cirasuolo}, {Garn}, \& {Dalton}}]{Sobral11}
{Sobral}, D., {Best}, P.~N., {Smail}, I., {et~al.} 2011, \mnras, 411, 675

\bibitem[{{Sobral} {et~al.}(2014){Sobral}, {Best}, {Smail}, {Mobasher},
  {Stott}, \& {Nisbet}}]{Sobral14}
---. 2014, \mnras, 437, 3516

\bibitem[{{Sobral} {et~al.}(2013){Sobral}, {Smail}, {Best}, {Geach}, {Matsuda},
  {Stott}, {Cirasuolo}, \& {Kurk}}]{Sobral13}
{Sobral}, D., {Smail}, I., {Best}, P.~N., {et~al.} 2013, \mnras, 428, 1128

\bibitem[{{Sobral} {et~al.}(2015){Sobral}, {Stroe}, {Dawson}, {Wittman}, {Jee},
  {R{\"o}ttgering}, {van Weeren}, \& {Br{\"u}ggen}}]{Sobral15}
{Sobral}, D., {Stroe}, A., {Dawson}, W.~A., {et~al.} 2015, \mnras, 450, 630

\bibitem[{{Sobral} {et~al.}(2016){Sobral}, {Stroe}, {Koyama}, {Darvish},
  {Calhau}, {Afonso}, {Kodama}, \& {Nakata}}]{Sobral16}
{Sobral}, D., {Stroe}, A., {Koyama}, Y., {et~al.} 2016, \mnras, 458, 3443

\bibitem[{{Stott} {et~al.}(2007){Stott}, {Smail}, {Edge}, {Ebeling}, {Smith},
  {Kneib}, \& {Pimbblet}}]{Stott07}
{Stott}, J.~P., {Smail}, I., {Edge}, A.~C., {et~al.} 2007, \apj, 661, 95

\bibitem[{{Strazzullo} {et~al.}(2013){Strazzullo}, {Gobat}, {Daddi}, {Onodera},
  {Carollo}, {Dickinson}, {Renzini}, {Arimoto}, {Cimatti}, {Finoguenov}, \&
  {Chary}}]{Strazzullo13}
{Strazzullo}, V., {Gobat}, R., {Daddi}, E., {et~al.} 2013, \apj, 772, 118

\bibitem[{{Stroe} {et~al.}(2015){Stroe}, {Sobral}, {Dawson}, {Jee}, {Hoekstra},
  {Wittman}, {van Weeren}, {Br{\"u}ggen}, \& {R{\"o}ttgering}}]{Stroe15}
{Stroe}, A., {Sobral}, D., {Dawson}, W., {et~al.} 2015, \mnras, 450, 646

\bibitem[{{Tran} {et~al.}(2010){Tran}, {Papovich}, {Saintonge}, {Brodwin},
  {Dunlop}, {Farrah}, {Finkelstein}, {Finkelstein}, {Lotz}, {McLure},
  {Momcheva}, \& {Willmer}}]{Tran10}
{Tran}, K.-V.~H., {Papovich}, C., {Saintonge}, A., {et~al.} 2010, \apjl, 719,
  L126

\bibitem[{{Tran} {et~al.}(2015){Tran}, {Nanayakkara}, {Yuan}, {Kacprzak},
  {Glazebrook}, {Kewley}, {Momcheva}, {Papovich}, {Quadri}, {Rudnick},
  {Saintonge}, {Spitler}, {Straatman}, \& {Tomczak}}]{Tran15}
{Tran}, K.-V.~H., {Nanayakkara}, T., {Yuan}, T., {et~al.} 2015, \apj, 811, 28

\bibitem[{{van den Bosch} {et~al.}(2008){van den Bosch}, {Aquino}, {Yang},
  {Mo}, {Pasquali}, {McIntosh}, {Weinmann}, \& {Kang}}]{Vandenbosch08}
{van den Bosch}, F.~C., {Aquino}, D., {Yang}, X., {et~al.} 2008, \mnras, 387,
  79

\bibitem[{{van der Wel} {et~al.}(2008){van der Wel}, {Holden}, {Zirm}, {Franx},
  {Rettura}, {Illingworth}, \& {Ford}}]{VanDerWel08}
{van der Wel}, A., {Holden}, B.~P., {Zirm}, A.~W., {et~al.} 2008, \apj, 688, 48

\bibitem[{{van der Wel} {et~al.}(2009){van der Wel}, {Rix}, {Holden}, {Bell},
  \& {Robaina}}]{VanDerWel09}
{van der Wel}, A., {Rix}, H.-W., {Holden}, B.~P., {Bell}, E.~F., \& {Robaina},
  A.~R. 2009, \apjl, 706, L120

\bibitem[{{van Dokkum} {et~al.}(2009){van Dokkum}, {Kriek}, \&
  {Franx}}]{vandokkum09}
{van Dokkum}, P.~G., {Kriek}, M., \& {Franx}, M. 2009, \nat, 460, 717

\bibitem[{{Vogelsberger} {et~al.}(2014){Vogelsberger}, {Genel}, {Springel},
  {Torrey}, {Sijacki}, {Xu}, {Snyder}, {Bird}, {Nelson}, \&
  {Hernquist}}]{Vogelsberger14}
{Vogelsberger}, M., {Genel}, S., {Springel}, V., {et~al.} 2014, \nat, 509, 177

\bibitem[{{von der Linden} {et~al.}(2010){von der Linden}, {Wild}, {Kauffmann},
  {White}, \& {Weinmann}}]{Vonderlinden10}
{von der Linden}, A., {Wild}, V., {Kauffmann}, G., {White}, S.~D.~M., \&
  {Weinmann}, S. 2010, \mnras, 404, 1231

\bibitem[{{Vulcani} {et~al.}(2010){Vulcani}, {Poggianti}, {Finn}, {Rudnick},
  {Desai}, \& {Bamford}}]{Vulcani10}
{Vulcani}, B., {Poggianti}, B.~M., {Finn}, R.~A., {et~al.} 2010, \apjl, 710, L1

\bibitem[{{Welikala} {et~al.}(2016){Welikala}, {B{\'e}thermin}, {Guery},
  {Strandet}, {Aird}, {Aravena}, {Ashby}, {Bothwell}, {Beelen}, {Bleem}, {de
  Breuck}, {Brodwin}, {Carlstrom}, {Chapman}, {Crawford}, {Dole}, {Dor{\'e}},
  {Everett}, {Flores-Cacho}, {Gonzalez}, {Gonz{\'a}lez-Nuevo}, {Greve},
  {Gullberg}, {Hezaveh}, {Holder}, {Holzapfel}, {Keisler}, {Lagache}, {Ma},
  {Malkan}, {Marrone}, {Mocanu}, {Montier}, {Murphy}, {Nesvadba}, {Omont},
  {Pointecouteau}, {Puget}, {Reichardt}, {Rotermund}, {Scott}, {Serra},
  {Spilker}, {Stalder}, {Stark}, {Story}, {Vanderlinde}, {Vieira}, \&
  {Wei{\ss}}}]{Welikala16}
{Welikala}, N., {B{\'e}thermin}, M., {Guery}, D., {et~al.} 2016, \mnras, 455,
  1629

\bibitem[{{Wetzel} {et~al.}(2013){Wetzel}, {Tinker}, {Conroy}, \& {van den
  Bosch}}]{Wetzel13}
{Wetzel}, A.~R., {Tinker}, J.~L., {Conroy}, C., \& {van den Bosch}, F.~C. 2013,
  \mnras, 432, 336

\bibitem[{{Wijesinghe} {et~al.}(2012){Wijesinghe}, {Hopkins}, {Brough},
  {Taylor}, {Norberg}, {Bauer}, {Brown}, {Cameron}, {Conselice}, {Croom},
  {Driver}, {Grootes}, {Jones}, {Kelvin}, {Loveday}, {Pimbblet}, {Popescu},
  {Prescott}, {Sharp}, {Baldry}, {Sadler}, {Liske}, {Robotham}, {Bamford},
  {Bland-Hawthorn}, {Gunawardhana}, {Meyer}, {Parkinson}, {Drinkwater},
  {Peacock}, \& {Tuffs}}]{Wijesinghe12}
{Wijesinghe}, D.~B., {Hopkins}, A.~M., {Brough}, S., {et~al.} 2012, \mnras,
  423, 3679

\bibitem[{{Xu} {et~al.}(2012){Xu}, {Zhao}, {Scoville}, {Capak}, {Drory}, \&
  {Gao}}]{Xu12}
{Xu}, C.~K., {Zhao}, Y., {Scoville}, N., {et~al.} 2012, \apj, 747, 85

\bibitem[{{Ziparo} {et~al.}(2014){Ziparo}, {Popesso}, {Finoguenov}, {Biviano},
  {Wuyts}, {Wilman}, {Salvato}, {Tanaka}, {Nandra}, {Lutz}, {Elbaz},
  {Dickinson}, {Altieri}, {Aussel}, {Berta}, {Cimatti}, {Fadda}, {Genzel}, {Le
  Floc'h}, {Magnelli}, {Nordon}, {Poglitsch}, {Pozzi}, {Portal}, {Tacconi},
  {Bauer}, {Brandt}, {Cappelluti}, {Cooper}, \& {Mulchaey}}]{Ziparo14}
{Ziparo}, F., {Popesso}, P., {Finoguenov}, A., {et~al.} 2014, \mnras, 437, 458

\end{thebibliography}

\appendix

\begin{figure*} 
 \begin{center}
\includegraphics[width=7in]{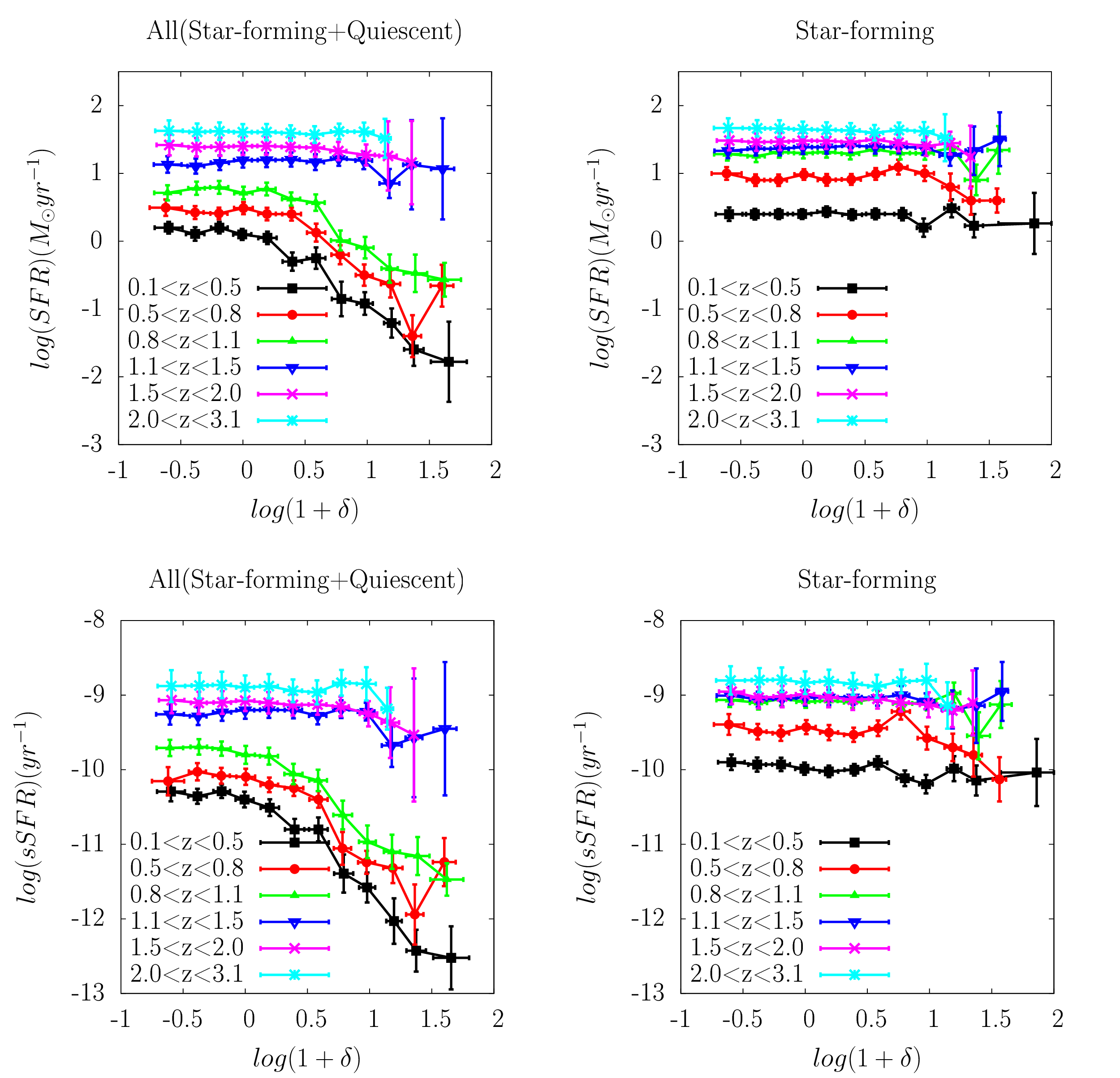}
\caption{Similar to Figure \ref{fig:SFR-sSFR} but based on a new sample selection. The new sample selection only comprises galaxies that are more massive than the mass completeness limit of the highest-$z$ sample (log($M/M_{\odot}$)$\geqslant$ 9.97). We find that our results presented in Section \ref{SFR-sSFR-env} do not depend on the sample selection.}
\label{fig:SFR-sSFR1}
 \end{center}
\end{figure*}

In Section \ref{SFR-sSFR-env}, we showed that at $z\lesssim$ 1, the median SFR and sSFR of galaxies strongly depend on their host environment. However, at $z\gtrsim$ 1, we found an environmental independence of the median SFR and sSFR. If the environment is mostly relevant for quenching the less massive galaxies, since our high-$z$ samples do not contain the less massive systems, the environmental invariance of the median SFR at $z\gtrsim$ 1 might be due to our selection of only the massive galaxies at higher redshifts. We investigate this by selecting only galaxies that are more massive than the mass completeness limit of our highest-$z$ sample (log($M/M_{\odot}$)$\geqslant$ 9.97) at all redshifts. Figure \ref{fig:SFR-sSFR1} shows the results given the new sample selection. We clearly see that our results given in Section \ref{SFR-sSFR-env} still hold with the new sample selection. We note that the median sSFR for low-$z$ samples are shifted toward lower values because of the selection of more massive galaxies in the new sample selection process. The slight environmental dependence (only seen in very dense regions) of the star-forming galaxies for the 0.5$<$ $z$ $<$0.8 sample is not significant at p $<$ 0.05 level ($<$ 2$\sigma$).  

\end{document}